\documentclass{emulateapj}
\usepackage{graphics}
\usepackage{tabu}
\usepackage{natbib}
\usepackage{color}
\usepackage{rotating}
\usepackage{multirow}
\usepackage{footnote}
\usepackage{datetime}
\usepackage{amsmath}
\usepackage{mathrsfs}
\usepackage[normalem]{ulem}
\begin{document}

\title{Correlations Between Black Holes and Host Galaxies in the Illustris and IllustrisTNG Simulations}

\author{Yuan Li\altaffilmark{1,2} , Melanie Habouzit\altaffilmark{1}, Shy Genel\altaffilmark{1,3}, Rachel Somerville\altaffilmark{1}, 
Bryan A. Terrazas\altaffilmark{4}, Eric F. Bell\altaffilmark{4}, Annalisa Pillepich\altaffilmark{5}, Dylan Nelson\altaffilmark{6}, Rainer Weinberger\altaffilmark{7}, Vicente Rodriguez-Gomez\altaffilmark{8}, Chung-Pei Ma\altaffilmark{2}, Ruediger Pakmor\altaffilmark{6}, Lars Hernquist\altaffilmark{7}, Mark Vogelsberger\altaffilmark{9}} 

\altaffiltext{$\dagger$}{yuan.astro@berkeley.edu}
\altaffiltext{1}{Center for Computational Astrophysics, Flatiron Institute, 162 5th Ave, New York, NY 10010, USA}
\altaffiltext{2}{Department of Astronomy, University of California, Berkeley, CA 94720, USA}
\altaffiltext{3}{Columbia Astrophysics Laboratory, Columbia University, 550 West 120th Street, New York, NY 10027, USA} 
\altaffiltext{4}{Department of Astronomy, University of Michigan, Ann Arbor, MI 48109, USA}
\altaffiltext{5}{Max-Planck-Institut f\"{u}r Astronomie, K\"{o}nigstuhl 17, D-69117 Heidelberg, Germany}
\altaffiltext{6}{Max Planck Institute for Astrophysics, Karl-Schwarzschild-Str. 1, 85741 Garching bei Munchen, Germany}
\altaffiltext{7}{Institute for Theory and Computation, Harvard-Smithsonian Center for Astrophysics, 60 Garden Street, Cambridge, MA 02138, USA}
\altaffiltext{8}{Instituto de Radioastronom\'ia y Astrof\'isica, Universidad Nacional Aut\'onoma de M\'exico, Apdo. Postal 72-3, 58089 Morelia, Mexico}
\altaffiltext{9}{Department of Physics, Kavli Institute for Astrophysics and Space Research, Massachusetts Institute of Technology, Cambridge, MA 02139, USA}

\begin{abstract}
{We study black hole - host galaxy correlations, and the relation between the over-massiveness (the distance from the average $M_{BH}-\sigma$ relation) of super-massive black holes (SMBHs) and star formation histories of their host galaxies in the Illustris and TNG100 simulations. We find that both simulations are able to produce black hole scaling relations in general agreement with observations at $z=0$, but with noticeable discrepancies. Both simulations show an offset from the observations for the $M_{BH}-\sigma$ relation, and the relation between $M_{BH}$ and the Sersic index. The relation between $M_{BH}$ and stellar mass $M_*$ is tighter than the observations, especially for TNG100. For massive galaxies in both simulations, the hosts of over-massive SMBHs (those above the mean $M_{BH}-\sigma$ relation) tend to have larger Sersic indices and lower baryon conversion efficiency, suggesting a multidimensional link between SMBHs and properties of their hosts. In Illustris, the hosts of over-massive SMBHs have formed earlier and have lower present-day star formation rates, in qualitative agreement with the observations for massive galaxies with $\sigma>100 \rm km/s$. For low-mass galaxies, such a correlation still holds in Illustris but does not exist in the observed data. For TNG100, the correlation between SMBH over-massiveness and star formation history is much weaker. The hosts of over-massive SMBHs generally have consistently larger star formation rates throughout history. These galaxies have higher stellar mass as well, due to the strong $M_{BH}-M_*$ correlation. Our findings show that simulated SMBH scaling relations and correlations are sensitive to features in the modeling of SMBHs.}

\end{abstract}

\section{Introduction}
\label{sec:intro} 
\setcounter{footnote}{0} 
It is widely accepted that the growth of SMBHs is linked to the evolution of their host galaxies, and that SMBHs also influence the hosts via active galactic nuclei (AGN) feedback \cite[see review by][and references therein]{McNamara2007, Fabian2012, Kormendy2013}. Depending on the level of accretion \citep{Churazov2005}, the forms of AGN feedback can be radiation, winds, jets, or a combination of them. 

At high accretion rates, the accretion disk around the SMBH is radiatively efficient, and the feedback is in the so-called quasar mode or radiative mode. Observational evidence for quasar mode feedback includes broad absorption line (BAL) winds seen close to the nuclei \citep{Chartas2003, Moe2009, Tombesi2015, Jiang2018}, and outflows on kpc or even larger galactic scales \citep{Dunn2010, CanoDaz2012, Arav2013, Carniani2015, Choi2015, Feruglio2015}. The main wind driving mechanism is still under debate, but many ideas have been explored from small-scale to cosmological-zoom simulations \citep{Proga2007, Hopkins2010, Choi2012, Liu2013, Costa2018}. In addition to winds, the radiation itself also has an impact on the surrounding gas \citep{Park2012, Qiu2018}.

At low accretion rates, the accretion disk around the SMBH is radiatively inefficient \citep{Yuan2014}, and feedback from SMBHs is thought to be mostly related to the relativistic jets \citep{Blandford2018}, but radiation can still have an effect \citep{Xie2017}. These SMBHs are often observed as bright radio sources. Thus this mode of feedback is often called radio mode feedback. Radio mode feedback is often observed in the center of cool-core galaxy clusters and massive galaxies, and is thought to be the key solution to the cooling flow problem \citep{Fabian1994}. In other words, radio-mode feedback is generally considered responsible for maintaining the quiescent state of massive galaxies in today's universe \citep{Schawinski2007}. Thus radio-mode feedback is also referred to as the maintenance mode. The effect of radio-mode feedback is well observed in nearby galaxy clusters and group centrals, in the form of X-ray ``bubbles'' (cavities) often filled with radio-emitting plasma, and ``ripples'' that are interpreted as shock waves or sound waves \citep[e.g.,][]{Fabian2006, Wise2007, Baldi2009, Blanton2011}. The energy associated with the cavities and waves is usually enough to offset radiative cooling in these systems \citep{Dunn2006, Rafferty2006}. Optical observations have also detected kinematic signatures of fast winds from host galaxies of low luminosity SMBHs \citep{Cheung2016, Penny2018}. Radio-mode feedback is often modeled as kinetic energy injection in numerical simulations \citep[e.g.,][]{Dubois2010, PII, Martizzi2019}. Hence it is also referred to as kinetic-mode feedback, even though quasar winds also carry kinetic energy, and some quasars (radio-loud quasars) are thought to produce jets too \citep{Wilson1995, Kellermann2016}. 

The physical process of AGN feedback is rather complicated, but small-scale general-relativistic magnetohydrodynamic simulations have made remarkable progress in the past few years on both accretion disk physics and the launching of the jets/winds \citep{Fragile2007, Tchekhovskoy2011, Bu2016, Jiang2017}. Galactic-scale and cosmological-zoom simulations have also explored different ways of implementing AGN feedback, studied the effects of different feedback mechanisms, and improved our understanding of the complex interplay between galactic weather, star formation, and black hole activities \citep{Omma2004, Fulai2011, Choi2012, Li2015, Yang2016, Yuan2018, Qiu2018}.

Another line of evidence supporting the idea of co-evolving galaxies and SMBHs is the observed black hole - host galaxy correlations. Among all black hole scaling relations, the most famous is the tight correlation between the mass of the SMBH $M_{BH}$ and the stellar velocity dispersion $\sigma$ of the bulge of the host galaxy \citep{Ferrarese2000, Gebhardt2000, Greene2010, McConnell2013, Kormendy2013, Woo2013}. However, $M_{BH}-\sigma$ is not the only tight scaling relation. $M_{BH}$ is also found to correlate with the luminosity and mass of the bulge \citep{Kormendy1993, Magorrian1998, Kayhan2009}. \citet{Graham2001} find that $M_{BH}$ also positively correlates with the concentration of the bulge, and the scatter is comparable to $M_{BH}-\sigma$ or even smaller. $M_{BH}$ also correlates with the total stellar mass of the host galaxy, but the scatter is much larger than $M_{BH} - M_{bulge}$ \citep{Reines2015, Terrazas2016, Shankar2019}. 

The interpretation of black hole scaling relations has been a subject of debate, and may not require co-evolution at all according to the central limit theorem \citep{Hirschmann2010, Jahnke2011}. It has also been suggested that black hole - galaxy correlations can be achieved in simulations without AGN feedback or self-regulation \citep{Daniel2013}. 

Recent observational studies by \citet{Martin2016, Terrazas2016, Martin2018Nature} suggest that SMBHs are linked to the quiescent state of the host galaxies. That is, SMBHs above the mean scaling relations (\citealt{Terrazas2016} use $M_{BH}-M_*$ and \citealt{Martin2018Nature} use $M_{BH}-\sigma$) tend to reside in galaxies that are more ``quenched'' (with lower star formation rate). \citet{Martin2018} also point out that such a correlation does not exist for smaller galaxies, suggesting that AGN feedback is less important in smaller systems.

Large-scale cosmological simulations usually employ simplistic sub-grid models for AGN feeding and feedback due to the limits of resolution and computational expense. Nonetheless, today's state-of-the-art cosmological simulations are able to recover many observed galaxy correlations and properties, such as the present-day stellar mass function and galaxy color bimodality \citep{EAGLE2015,  Nelson2018, Romeel2019}. Many simulations have also been shown to successfully reproduce some of the BH-host galaxy scaling relations in good agreement with observations \citep{Sijacki2015, McAlpine2017, Weinberger2018}. The relationship between SMBHs and their host properties carries a lot of information, but it is not clear how those relationships inform us about the physical processes that connect SMBHs and their hosts.

In this paper, we use the Illustris \citep{Vogelsberger2014,Vogelsberger2014Nature, Genel2014} and TNG100 \citep{Springel2018, Pillepich2018TNG, Marinacci2018, Nelson2018, Naiman2018} simulations to study black hole scaling relations, and the relation between the over-massiveness of the SMBH and the star formation history of the host galaxy. The goal of our study is to better understand how simulated black hole scaling relations are related to sub-grid models of black hole physics, and to gain more insights into how SMBHs are related to the star formation history of their host galaxies. We connect the results with those determined by selected observational datasets, albeit taken at face value and without correcting for observational selection biases.

The paper is structured as follows: Section~\ref{sec:simulations} describes the key aspects of the Illustris and TNG100 simulations; in Section~\ref{sec:results1}, we present the black hole-host galaxy scaling relations in Illustris and TNG100; in Section~\ref{sec:results2}, we discuss how the the over-massiveness of the SMBHs is related to the quenching of their host galaxies, and compare what we find for Illustris and TNG100 with the findings in \citet{Martin2018Nature} and \citet{Martin2018}. We conclude this work in Section~\ref{sec:conclusions}. 

\section{Simulations}
\label{sec:simulations}
The Illustris and TNG100 simulations are large-scale cosmological simulations that include prescriptions for star formation, stellar evolution and feedback, black hole formation, growth and feedback, and other physical processes relevant to the formation and evolution of galaxies. Both simulations are performed using the moving mesh code AREPO \citep{Springel2010, Pakmor2016}. Illustris is an N-body/hydrodynamical simulation and TNG100 also includes magnetohydrodynamics. TNG100 is one of the flagship runs of the IllustrisTNG project, a successor to the original Illustris simulation, with an improved numerical scheme and updated sub-grid models. The most notable changes are the revised implementation of galactic winds (stellar feedback) and a new black hole feedback model at low accretion rates. More detailed descriptions of the sub-grid models can be found in \citet{Vogelsberger2013} for Illustris, and in \citet{Pillepich2018} and \citet{Weinberger2017,Weinberger2018} for the IllustrisTNG model adopted in TNG100. For the purpose of this work, we summarize the key aspects of the sub-grid models related to SMBHs used in Illustris and TNG100. 

Three important physical processes related to SMBHs are seeding, feeding, and feedback. The seeding of SMBHs in Illustris and TNG100 are similar, but the seed masses are quite different. In Illustris, a SMBH particle with a mass of $1.42 \times 10^5 M_{\odot}$ is seeded in all halos above $7.1\times 10^{10}  M_{\odot}$ that do not already contain a SMBH. In TNG100, halos above $7.38 \times 10^{10}  M_{\odot}$ are seeded with SMBHs of $1.18 \times 10^6  M_{\odot}$, almost an order of magnitude larger than the seed mass in Illustris. 

The accretion onto SMBHs (feeding) in both Illustris and TNG100 is the Eddington-limited Bondi accretion: $\rm \dot{M}_{SMBH}=min (\dot{M}_{Bondi},\dot{M}_{Edd})$. The details differ in two important aspects. First, Illustris uses a ``boosted Bondi rate'' with an artificial boosting factor $\alpha=100$, while TNG100 does not. From a practical point of view, boosted Bondi accretion is commonly used in cosmological simulations to grow SMBHs efficiently \citep[e.g.,][]{Springel2005, Khalatyan2008}. The physical explanation for a boosted Bondi rate is that cosmological simulations do not resolve Bondi radii, and the Bondi accretion rate computed at the actual Bondi radius is likely higher than that computed based on the gas properties at larger radii. Multiphase gas is also poorly resolved in cosmological simulations, and a boosting factor can account for the accretion of cold gas \citep{Booth2009}. TNG100 does not need the boosting factor due to the large SMBH seeds. The second difference in the BH growth scheme is that in Illustris the properties of the gas that determine the Bondi formula are estimated by using only the parent gas cell where the BH is positioned, while TNG100 uses a kernel-weighted average over neighboring cells.

Feedback from SMBHs is, in both cases, divided into two modes based on the accretion rate normalized by the Eddington rate (often referred to as the Eddington ratio): quasar mode and radio mode \citep{Churazov2005}. When the accretion rate is high, the radiative efficiency is high and the SMBH is in the quasar mode (sometimes also referred to as the ``radiative mode'' or ``thermal mode'' in the literature). 
In both Illustris and TNG100, this quasar mode feedback is modeled as a continuous injection of thermal energy into a number of surrounding cells \footnote{For llustris-1 and TNG100-1, the number of cells is 256. We use the highest resolution simulations with 100 Mpc boxes from both suites (Illustris-1 and TNG100-1) in this study.}. The energy injection rate is $\rm \dot{E}_{therm} = 0.01 \dot{M} c^2$ in Illustris and $\rm \dot{E}_{therm} = 0.02 \dot{M} c^2$ in TNG100. The feedback from SMBH at low accretion rates is often referred to as the radio mode (sometimes also referred to as ``maintenance'', ``jet'' or ``wind mode''), and is modeled very differently in Illustris and TNG100. 
Because there is no assumption of radio emission in either Illustris or TNG100, we refer to this mode as ``low state mode'' throughout the rest of the paper. 
Low state mode feedback in Illustris is modeled by injecting thermal energy into hot bubbles at some distance from the SMBH. In TNG100, low state mode feedback is modeled by injecting pulsed kinetic energy into a number of cells near the SMBHs in a randomly oriented direction. The kinetic luminosity is $\rm \dot{E}_{kinetic}=\dot{M} c^2 \times min(\rho/0.05\rho_{SF}, 0.2)$, where $\rm \rho_{SF}$ is the star formation threshold density. Another unique feature in TNG100 is that the dividing line between quasar mode and low state mode feedback is no longer a fixed fraction of the Eddington ratio, as is the case in Illustris and most of the other cosmological simulations. Instead, the dividing line is also a function of the black hole mass:
\begin{equation}\label{eq:BH}
\rm \chi = min\Big [ 0.002 \big (\frac{ \it M_{BH}}{10^8M_{\odot}} \big )^2, 0.1\Big] \, .
\end{equation}
As a result of this feature, SMBHs in TNG100 are mostly in the quasar (thermal) mode when their mass is much lower than $10^8 \rm M_\odot$, and transition to mostly low state (kinetic) mode at $\gtrsim 10^8 \rm M_\odot$. 

In both Illustris and TNG100, the implementation of SMBH feedback is inspired by the $M_{BH}-\sigma$ relation or previous studies on the relation \citep{Springel2005}. 
However, neither Illustris nor TNG100 is specifically tuned to reproduce exactly all known black hole - host galaxy correlations. The relation between BH mass and galaxy stellar mass \citep{Magorrian1998} has been taken into consideration in both cases during the model development, but no exact value of the model parameters had been chosen so that the model outcome could fit precisely any specific relation (e.g., the \citet{Kormendy2013} relation). Thus all the correlations studied in this work can be considered as an emerging outcome of the simulations.

IllustrisTNG (www.tng-project.org) is a suite of simulations that have boxes of different sizes and resolutions. For this work, we exclusively work with the highest resolution realization of a $\sim 100^3$ co-moving $\rm Mpc^3$ box. We make the same choice for Illustris: in fact, TNG100 and Illustris have the same resolution and initial conditions, and are both evolved from $z=127$ to $z=0$. 
For Illustris, all the analyses are limited to galaxies with a stellar mass larger than $\sim 10^{10} M_{\odot}$ (the details of the catalog can be found in \citet{Xu2017}). For TNG100, the selection criterion is stellar masses larger than $\sim 5\times10^9 M_{\odot}$ within central 30 kpc (using the catalog from \citet{Xu2019}). The total number of galaxies in the Illustris and TNG100 catalogs are 6808 and 9686, respectively. The catalogs in \citet{Xu2017} and \citet{Xu2019} were originally made for different observational comparison purposes, and thus have different mass cuts. We have experimented with the same mass cut at $10^{10} M_{\odot}$ for TNG100. This does not change our main findings, but cuts out a significant fraction of galaxies with low BH masses. Because TNG100 galaxies tend to have slightly lower stellar masses on average, a slightly lower mass cut for TNG100 actually produces a more equivalent sample. Thus we decide to keep the original sample with different mass cuts for the two simulations.

The measurements of the physical quantities of simulated galaxies are summarized in Table~\ref{table:data}. More details are discussed in the rest of the paper.

\begin{table*}[]
\caption{Summary of analysis methods and measurements of galaxy properties. }
\label{table:data}
\begin{center}
\begin{tabular}{|p{3cm}|p{6.5cm}|p{7cm}|}
\hline
properties & observations & simulations in this work \\ \hline
$M_{BH}$ & \citet{Kormendy2013, McConnell2013, Martin2018Nature}  &sum of the masses of all blackholes in the halo \\ \hline
$M_*$ & \citet{Reines2015, Terrazas2016} & sum of all star particle masses within twice the stellar half mass radius \\ \hline
$\sigma$ &\citet{Kormendy2013, McConnell2013, Martin2018Nature}  & rest-frame SDSS-r band luminosity-weighted stellar line-of-sight velocity dispersion measured within a projected radius of 1.5 arcsec from galaxy center in x-projection \citep{Xu2017,Xu2019} \\ \hline
Sersic index & \citet{Graham2007, Savorgnan2016, Davis2019} & synthetic images designed to match Pan-STARRS observations \citep{Rodriguez-Gomez2019}.  \\ \hline
star formation history  & measured by \citet{Martin2018Nature} using the STECKMAP code \citep{Ocvirk2006} with the MILES \citep{Vazdekis2010} stellar population synthesis models  &  computed based on the formation histories of all the star particles that reside within the galaxy at z = 0. We use 0.5 $R_e$ for massive galaxies and 4.5 kpc for low-mass galaxies. \\ \hline
\end{tabular}\\ \ \\
\end{center}
\end{table*}

\section{SMBH - host galaxy correlations}\label{sec:results1}

\begin{figure*}
\begin{center}
\includegraphics[scale=0.38,trim=2.2cm 1cm 2.5cm 1cm, clip=true]{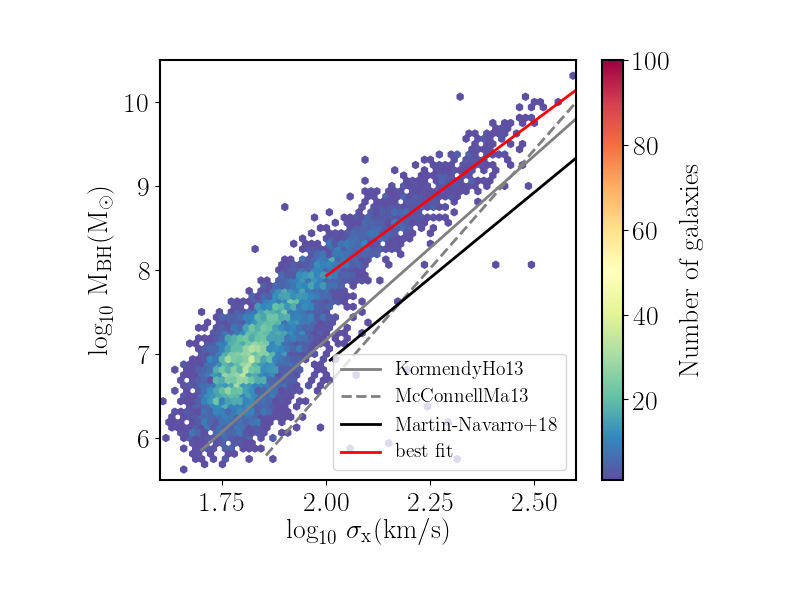}
\includegraphics[scale=0.38,trim=2.2cm 1cm 2.5cm 1cm, clip=true]{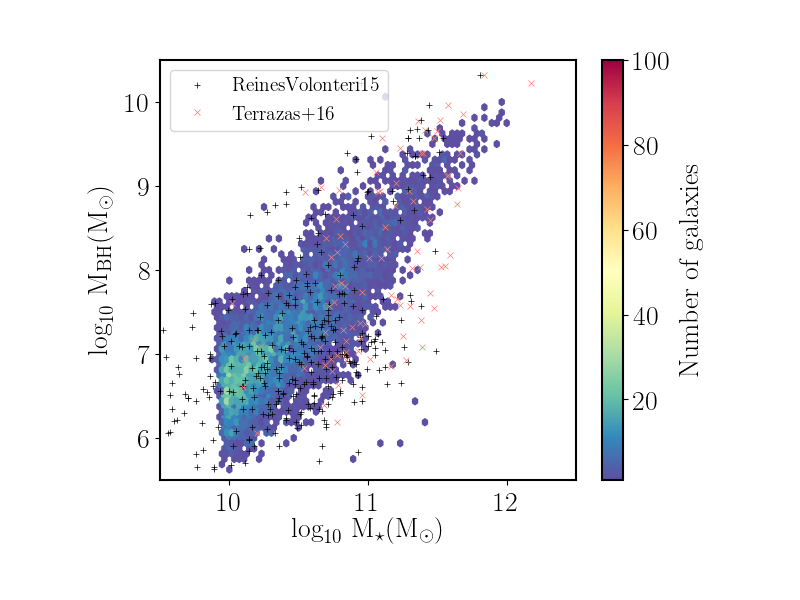}
\includegraphics[scale=0.38,trim=2.2cm 1cm 2.5cm 1cm, clip=true]{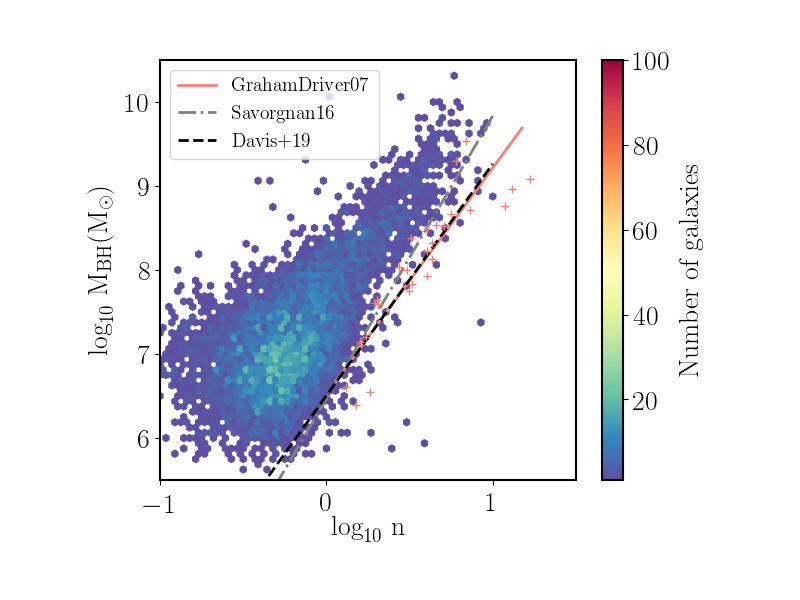}

\includegraphics[scale=0.38,trim=2.2cm 1cm 2.5cm 1cm, clip=true]{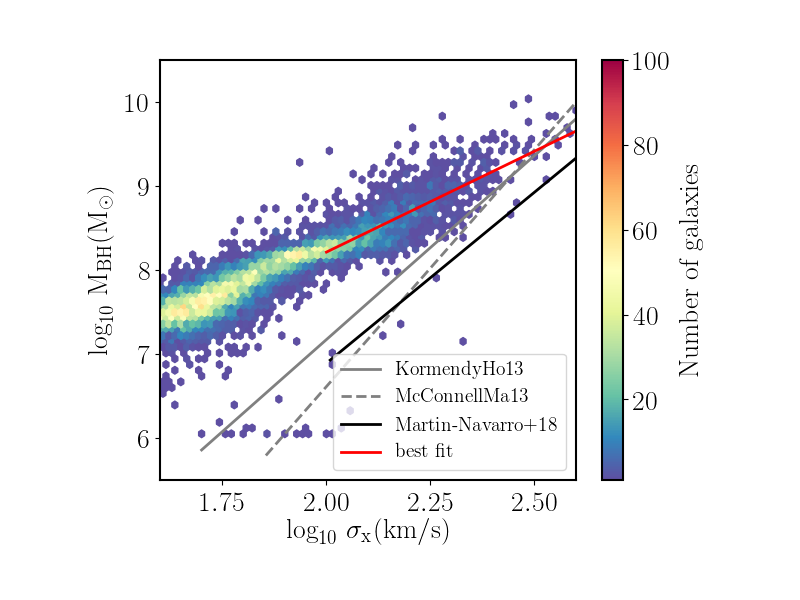}
\includegraphics[scale=0.38,trim=2.2cm 1cm 2.5cm 1cm, clip=true]{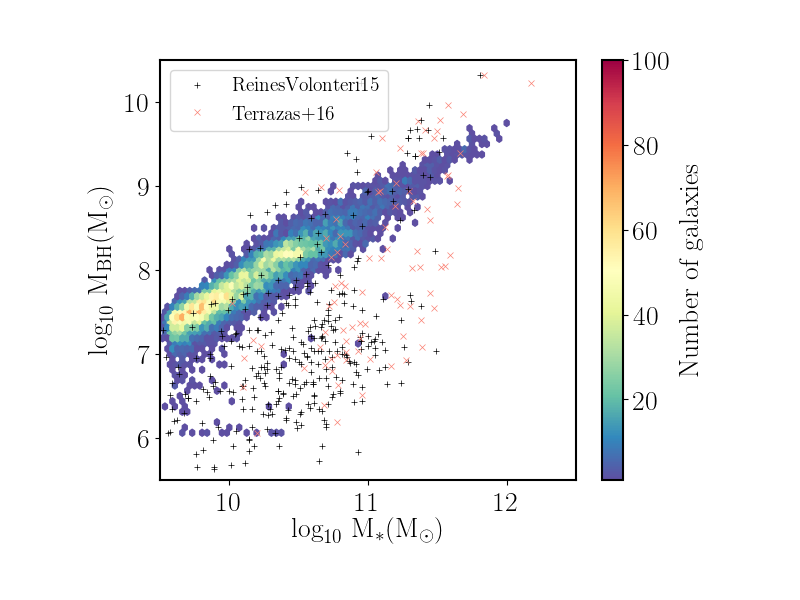}
\includegraphics[scale=0.38,trim=2.2cm 1cm 2.5cm 1cm, clip=true]{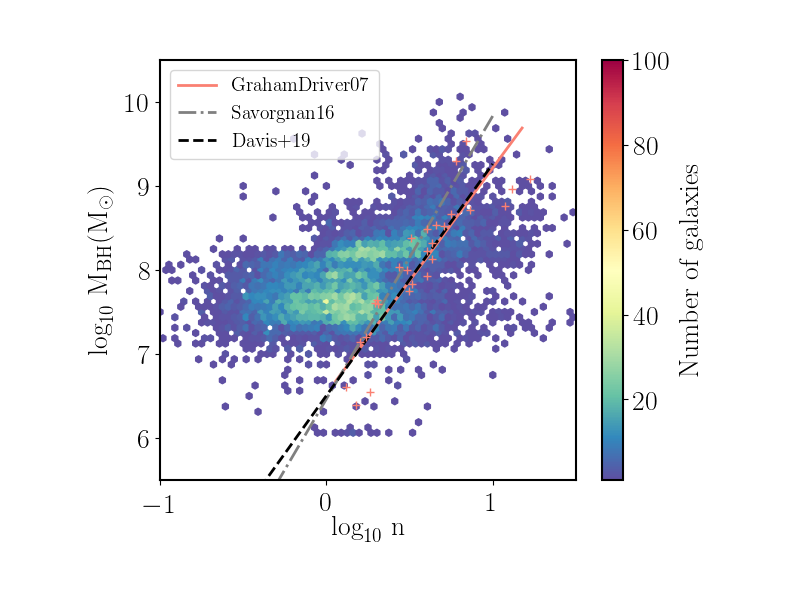}

\caption{SMBH - host galaxy correlations in Illustris (top) and TNG100 (bottom) simulations. The left panels are showing the $M_{BH}-\sigma$ relation compared with the observed best-fitting relations \citep{Kormendy2013, McConnell2013, Martin2018Nature}. The middle panels are showing the $M_{BH}-M_*$ relation compared with the observed data in \citet{Reines2015} and \citet{Terrazas2016}. The right panels show the correlation between $M_{BH}$ and Sersic index compared with the linear best-fitting relations from \citet{Graham2007, Savorgnan2016, Davis2019} for the observed galaxies. Also plotted are the data points from \citet{Graham2007}. For clarify, error bars of the observed data are not plotted here. Simulation data only include galaxies with $M_*\gtrsim10^{10}M_{\odot}$ for Illustris and $M_*\gtrsim 5\times10^9M_{\odot}$ for TNG100. 
\label{fig:M_sigma}}
\end{center}
\end{figure*}

\begin{figure}
\begin{center}
\includegraphics[scale=0.36,trim=0cm 0cm 0cm 0cm, clip=true]{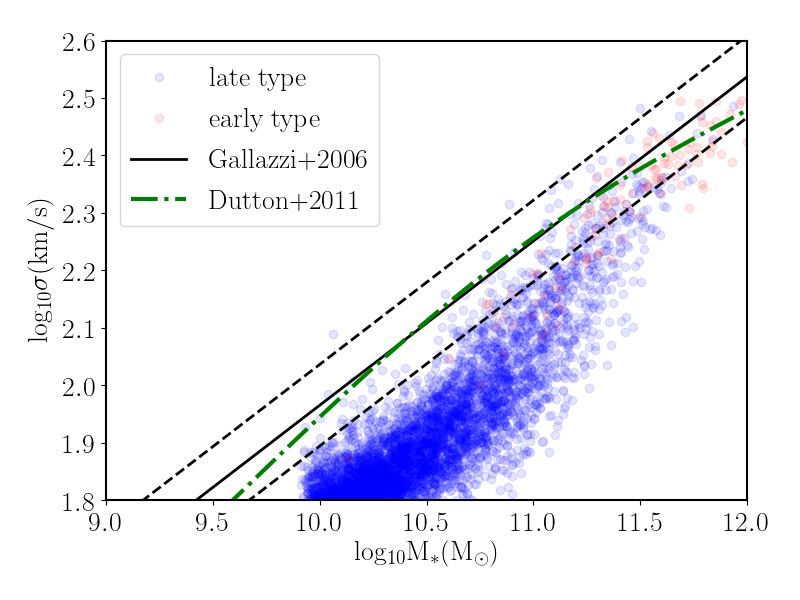}
\includegraphics[scale=0.36,trim=0cm 0cm 0cm 0cm, clip=true]{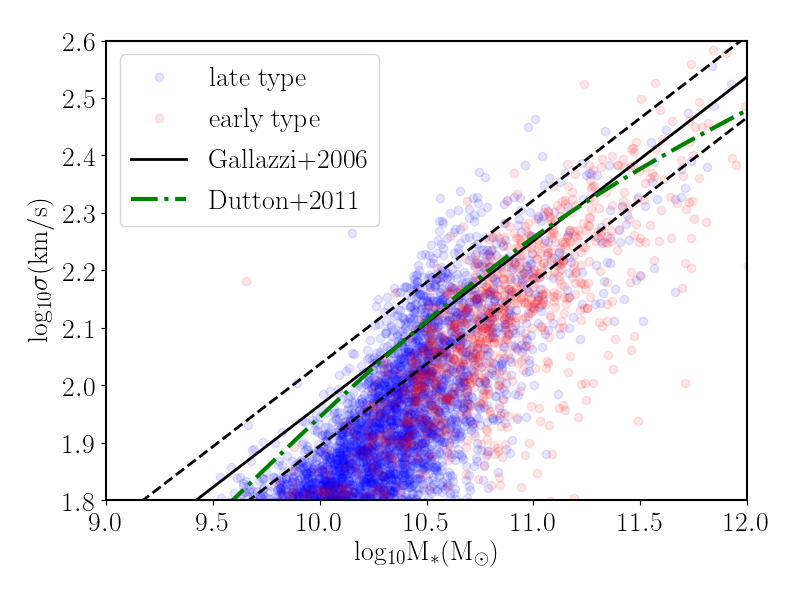}

\caption{The relation between stellar velocity dispersion $\sigma$ and stellar mass $M_*$ for galaxies in Illustris (top) and TNG100 (bottom). Blue dots show late type galaxies (defined as those with Sersic indices smaller than 4) and red dots show early type galaxies (Sersic indices $\geq$ 4). The solid black line is the linear fit to the Faber Jackson Relation in \citet{Gallazzi2006} for early-type galaxies with $R_{90}/R_{50}>2.8$ in the Sloan Digital Sky Survey (SDSS) Data Release 2 \citep{Abazajian2004} and dashed lines show the scatter. The dotted dash green line shows the best fit for the early-type galaxies in SDSS Data Release 7 \citep{Abazajian2009} using a double power law \citep{Dutton2011}. For given $M_*$, $\sigma$ tend to be low in simulations, and the discrepancy is larger in Illustris. Massive galaxies in TNG100 agree reasonably well with the observations, although with a tendency towards low $\sigma$ at a given $M_*$.
\label{fig:FJR}}
\end{center}
\end{figure}

In this section, we discuss the black hole - host galaxy correlations in both Illustris and TNG100.

Figure~\ref{fig:M_sigma} shows the $M_{BH}-\sigma$ relation (left panels), $M_{BH}-M_*$ relation (middle panels), and the relation between $M_{BH}$ and the Sersic index $n$ of the host galaxy (right panels) for both Illustris (top) and TNG100 (bottom). 

\subsection{The $M_{BH}-\sigma$ relation}\label{sec:Msigma}

For comparison, we have plotted the best fitting $M_{BH}-\sigma$ relations from the observations by \citet{Kormendy2013} and \citet{McConnell2013} for all galaxies, and \citet{Martin2018Nature} for the massive galaxies with $\sigma >100 \rm km/s$\footnote{In this paper, we generally refer to galaxies with $\sigma >100 \rm km/s$ as massive galaxies, and galaxies with $\sigma <100 \rm km/s$ as low-mass galaxies.} in the left panels of Figure~\ref{fig:M_sigma}. The sample in \citet{Martin2018Nature} is selected from \citet{van2016}. The massive galaxies in Illustris lie above the observed $M_{BH}-\sigma$ relations, but the slope is roughly consistent. At the low-mass end, $M_{BH}-\sigma$ in Illustris bends down and becomes closer to the observations, although \citet{Martin2018} suggests that the observed relation bends up (the slope is shallower) for the low-mass galaxies. The massive galaxies in TNG100 are also above the observations (except for some of the very massive ones). The linear best fit (the red line) has a slope that is smaller than the observed slope, but it is also obvious that the $M_{BH}-\sigma$ relation in TNG100 cannot be well described with a linear fit. The relation appears to have a break at $\rm log_{10}\sigma (km/s)\sim 2.1$, and from there to $\rm log_{10}\sigma (km/s) \sim 1.9$, the relation flattens. This also corresponds to a black hole mass of $\gtrsim 10^8M_\odot$, which is the critical mass that helps determine feedback modes, as described in Section~\ref{sec:simulations}. The relation steepens again at $\rm log_{10}\sigma\rm(km/s) < 1.9$, but quickly flattens at the low $\sigma$ end to $\rm log_{10} M_{BH}(M_\odot)\sim7.5$. 

It is unlikely that the discrepancy between simulations and observations can be attributed to the different ways galaxy properties are measured. For both Illustris and TNG100, we use $\sigma_x$, which is the rest-frame SDSS-r band luminosity-weighted stellar line-of-sight velocity dispersion measured within a projected radius of 1.5 arcsec from galaxy centre in x-projection \citep{Xu2017,Xu2019}. Since there is no preferred direction in Illustris or TNG100, projections along y- and z-axis are the same and we have verified this. We have also tried to use $\sigma$ measured within 0.5 $R_e$ and 2.0 $R_e$, and found very little change. If we use the stellar-mass-weighted $\sigma$ instead of the light-weighted $\sigma$, the $M_{BH}-\sigma$ relation steepens very slightly, but the general results are consistent. The Illustris $M_{BH}-\sigma$ also appears consistent with the relation presented in \citet{Sijacki2015} which independently computed $\sigma$ directly from the simulation data. 

The main discrepancy between simulations and observations for both Illustris and TNG100 is that on the $M_{BH}-\sigma$ diagram, the simulated galaxies are above the observed $M_{BH}-\sigma$ relation (Figure~\ref{fig:M_sigma}). This is possibly due to SMBHs being overly massive in simulations and/or $\sigma$ being too low. The black hole mass function in Illustris is in good agreement with the observations \citep{Sijacki2015} and the TNG100 quasar luminosity function at low redshift also agrees well with the observations \citep{Weinberger2018, Habouzit2018}. This suggests that at least in the local universe, at the high-mass end, SMBHs in these simulations are generally not too massive. Thus, overly massive SMBHs cannot be the main reason for the order of magnitude offset in the $M_{BH}-\sigma$ relation. 

At the high-mass end, the discrepancy between the $M_{BH}-\sigma$ relations for Illustris and massive galaxies in TNG100 likely has to do with $\sigma$ being too small in simulations. Figure~\ref{fig:FJR} shows the relation between $\sigma$ and total stellar mass of a galaxy (Faber Jackson Relation) in Illustris (top) and TNG100 (bottom). We also separate early type and late type galaxies based on their Sersic indices. Strictly, the Faber Jackson Relation only applies to early type galaxies. Because there are very few galaxies with Sersic index $>4$ in Illustris, and because simulations may under-predict Sersic indices, we show both early and late type galaxies in Figure~\ref{fig:FJR}. In both simulations, for a given $M_*$, $\sigma$ tends to be lower than the observed relation, and the disagreement is worse for lower mass galaxies. The best agreement is found between observations and massive galaxies in TNG100. The Illustris Faber Jackson Relation is discussed in more detail in \citet{Xu2017}, which also shows that the sizes of the Illustris galaxies tend to be too large. For massive galaxies in TNG100, \citet{Genel2018} find that the effective radius is overestimated by 0.1-0.15 dex at $M_* > 10^{10.5} M_\odot$ \citep[or up to 1.5 times, see also][]{Wang2018, Rodriguez-Gomez2019}. For a given mass, $\sigma$ and $R_{e}$ are roughly related to each other as $\sigma^2\sim1/R_{e}$. Thus a 0.1-0.2 dex difference in $R_{e}$ corresponds to a 0.05-0.1 dex difference in $\rm log_{10}\sigma (km/s)$, and can explain the discrepancy seen in Figure~\ref{fig:M_sigma} for massive galaxies. The better agreement in Faber Jackson Relation for TNG100 galaxies with $\rm log_{10}\sigma(km/s) > 2.2$ is also consistent with our finding that the massive galaxies in TNG100 have a better agreement in $M_{BH}-\sigma$ relation. In Illustris, the sizes of galaxies are even more over-estimated, $\sim$ twice as large as the observed galaxies on average for all the galaxies we study here ($M_*>10^{10}M_\odot$) \citep{Snyder2015,Bottrell2017}. This is consistent with Figure~\ref{fig:FJR} that shows for a given $M_*$, the offset in $\sigma$ is larger for Illustris than for TNG100. A factor of 2 difference in $R_e$ corresponds to a 0.15 dex shift in $\rm log_{10}\sigma (km/s)$, which is again consistent with the offset we find in the Illustris $M_{BH}-\sigma$ relation in Figure~\ref{fig:M_sigma}.

For low-mass galaxies ($\rm log_{10}\sigma (km/s) < 2.0$), the galaxy sizes in TNG100 agree well with the observations \citep{Genel2018}, and the offset in $\sigma$ shown in Figure~\ref{fig:FJR} is not enough to explain the discrepancy we see in the TNG $M_{BH}-\sigma$ relation for low-mass galaxies. Therefore, in low-mass galaxies, SMBHs in TNG100 are likely too massive. Two factors may contribute to this discrepancy. First, SMBH seeds in TNG100 may be too massive $\sim 10^6 M_\odot$. Since Bondi accretion rate scales as $M^2_{BH}$, more massive seeds also lead to more accretion onto the SMBH. In addition, low-mass SMBHs in TNG100 tend to be in quasar mode, and the pure thermal quasar mode feedback in TNG100 is not very effective at removing gas from the central region of the host galaxy. Thus the growth of SMBHs in quasar mode may be too fast. The over-efficient early growth of SMBHs is also used to explain overly abundant luminous SMBHs at high redshift in TNG100 \citep{Weinberger2018}. 

\subsection{The $M_{BH}-M_*$ relation}

The middle panels of Figure~\ref{fig:M_sigma} show the $M_{BH}-M_*$ relation in Illustris (top) and TNG100 (bottom). The cut at low-mass end has to do with a cut in the catalog (see Section~\ref{sec:simulations} for details). For comparison, we have also over-plotted the observational data compiled in \citet{Terrazas2016} with dynamical estimates of BH masses, and \citet{Reines2015} which includes a sample of broad-line AGNs at low redshift, as well as galaxies with dynamical BH masses. The Illustris $M_{BH}-M_*$ relation compared with different observations has been shown in \citet{Sijacki2015} and \citet{Terrazas2016}; the TNG300 $M_{BH}-M_*$ relation is shown in \citet{Weinberger2018}, and is discussed in detail in a separate paper \citep{Terrazas2019}. Here we only focus on the key aspects that are relevant to the discussions in this work. 

Both Illustris and TNG100 show correlations between the total stellar mass of a galaxy $M_*$ and $M_{BH}$ that are very broadly in agreement with observations. However, the scatter is smaller than the observations. The means are offset between observations and models too. For instance, at $log_{10}M_*(\rm M_\odot)=10.2$, The mean $log_{10}M_{BH}(\rm M_\odot)$ in the \citet{Reines2015} sample is 6.7, with a $1\sigma$ scatter of 0.6. The mean $log_{10}M_{BH}(\rm M_\odot)$ in Illustris and TNG100 at $log_{10}M_*(\rm M_\odot)=10.2$ are 6.9 and 7.9, respectively, and the scatters are $\sim 0.4$ for Illustris and 0.2 for TNG100. Models hug the high BH mass envelope of galaxies, and do not produce the galaxies with lower BH mass at a given stellar mass. This is especially pronounced for TNG100, which produces an $M_{BH}-M_*$ correlation that is even tighter than the $M_{BH}-\sigma$ relation. 

Many factors may contribute to the difference in the amount of scatter between simulations and observations. First, due to resolution limit, simulations do not fully capture the stochasticity of black hole accretion and feedback. Second, the SMBH seeding scheme is such that the initial SMBH seed mass is tightly linked to the halo mass, which is correlated with the stellar mass. The central limit theorem \citep{Hirschmann2010, Jahnke2011} suggests that these correlations should get tighter through subsequent mergers. Third, the value of the radiative efficiency is set partly by the spin of the black hole and the geometry of the accretion flow \citep{Bustamante2019}, a dependence that is no not included in the the simulations. The same applies to the coupling efficiency, which is currently set to a fixed constant value in both simulations. It is plausible to believe that, if a wider range of distributions in these factors had been allowed in the models, the scatter in black hole mass at a given stellar mass could have been larger.
Additionally, in TNG100, the accretion rate onto SMBHs is computed using a kernel-weighted average over 256 neighboring cells. As a result, the SMBH accretion is correlated with the gas properties of a volume that makes up a non-negligible fraction of the star forming region of the host galaxy. This may be one reason why TNG100 produces an even tighter $M_{BH}-M_*$ correlation than Illustris. 

\subsection{The $M_{BH}-$ Sersic index relation}

The right panels of Figure~\ref{fig:M_sigma} show the relation between $M_{BH}$ and the Sersic index $n$ of the host galaxy which measures the concentration of light from the stars in the galaxy. 

{The Sersic indices are measured using synthetic images of simulated galaxies created with the SKIRT \citep{Baes2011} radiative transfer code which include the effects of dust attenuation and scattering, and are designed to match Pan-STARRS observations \citep{Rodriguez-Gomez2019}. For Illustris galaxies, we have compared our Sersic indices with those obtained by fitting a 2D Sersic profile to the radial distribution of the elliptical isophotes of mock SDSS images of simulated galaxies \citep{Xu2017}. For galaxies with low Sersic indices, the dust corrected Sersic indices tend to be slightly lower than the uncorrected measurements, but the overall agreement between the two measurements is very good.} 

Also plotted in the right panels of Figure~\ref{fig:M_sigma} is the best linear fit for the 27 observed galaxies in \citet{Graham2007} \footnote{Although a linear relation provides a good fit to the data, \citet{Graham2007} finds that the best-fitting quadratic relation has smaller scatter. We use the linear fit here for simplicity.}. The data in \citet{Graham2007} are from \citet{Graham2001} which use SMBHs from \citet{Msigma} with revised $M_{BH}$ estimates from \citet{Kormendy2001}, and Sersic indices based on high-quality R-band images. The actual data are plotted as pink symbols. We have also included the best linear fit lines from \citet{Savorgnan2016} and \citet{Davis2019}. \citet{Savorgnan2016} uses 66 galaxies with dynamical measurement of $M_{BH}$ and for which they are able to successfully model the light distribution and measure the spheroid structural parameters using $3.6\mu m$ Spitzer satellite images. \citet{Davis2019} measures $M_{BH}-n$ relation for a sample of 40 spiral galaxies using the spheroid major axis Sersic indices.

The $M_{BH}-n$ relations in both Illlustris and TNG100 show a trend in general agreement with the observations. The simulated data are above (or shifted to the left of) the observed relation. This shift is larger in Illustris than TNG100, especially at the high mass end. Similar to the offset in $M_{BH}-\sigma$ relation, we do not think that this discrepancy is solely due to SMBHs being overly massive in simulations. 

As \citet{Rodriguez-Gomez2019} shows by mocking simulated galaxies as if imaged within Pan-STARRS observations, the Sersic indices of Illustris galaxies are lower by a factor of 2-3 in comparison to Pan-STARRS. TNG100 galaxies have Sersic indices that are in much better agreement with the Pan-STARRS observations than Illustris, and are only slightly lower at $M_*\sim$ a few times $10^{10} M_\odot$. However, because of the small range of Sersic indices compared with the range of $M_{BH}$ which span orders of magnitude, a small mismatch in Sersic can result in a large apparent discrepancy in the $M_{BH}-n$ relation. According to \citet{Rodriguez-Gomez2019}, at $M_*$ of a few times $10^{10} M_\odot$, the average Sersic index of the Pan-STARRS galaxies is $\sim 3$, and the average Illustris and TNG100 Sersic indices for the same stellar mass is about $\sim 0.7$ and $1.5-2$, respectively. These correspond to a shift of $\sim0.6$ dex and $\sim 0.2-0.3$ dex in $log_{10}n$ for Illustris and TNG100, respectively, and are consistent with what we see in the right panel of Figure~\ref{fig:M_sigma}. Therefore, it is plausible that, if simulated galaxies had higher Sersic indices n (and thus in better agreement with Pan-STARRS data), they would better agree with the observed $M_{BH}-n$ relations in Figure~\ref{fig:M_sigma}.

It has previously been suggested that if simulated galaxies under predict $\sigma$ and Sersic n, this could be due to insufficient numerical resolution \citep[e.g., see][for discussions on how resolution affects the central density profiles of simulated galaxies using the FIRE-2 physical model]{Hopkins2018}. However, resolution cannot be the only issue here. Illustris and TNG100 galaxies are modeled at the same mass and spatial resolution, yet they exhibit different size-mass and Sersic index-mass relations (see \citet{Pillepich2018, Rodriguez-Gomez2019}). 

As is pointed out in \citet{Graham2001} and \citet{Graham2007}, $log_{10}M_{BH}$ and $log_{10}n$ may not be linearly correlated. While the $M_{BH}-n$ relation in Illustris could be well described with a linear relation with some scatter, this is certainly not the case for TNG100. In TNG100, there is a concentration of galaxies with $log_{10}n\sim0-0.3$ and $M_{BH}\sim$ a few times $\rm 10^7M_\odot$, and a horizontal strip at $M_{BH} \gtrsim 10^8 M_{\odot}$. These features are consistent with what is seen in the other black hole scaling relations, and are likely related to the large seed mass, the ineffective quasar mode feedback, and the critical $M_{BH}$ in the feedback mode transition discussed previously. 

\subsection{The importance of features in the black hole scaling relations}

As Figure~\ref{fig:M_sigma} shows, all the TNG100 black hole - host galaxy correlations show a horizontal feature at the critical $M_{BH}$ (it is less prominent in the $M_{BH}-M_*$ relation but still noticeable). The reason may be that SMBH feedback is not efficient enough below $10^8 M_{\odot}$, allowing fast growth for small SMBHs, but becomes slightly too efficient at a few times $10^8 M_{\odot}$ \citep[see also][]{Habouzit2018}, and starts to suppress the growth of the SMBHs themselves. As the galaxies continue to evolve, the growth of the SMBH lags behind the growth of other quantities such as $\sigma$ and $M_*$. This results in a pile-up of galaxies with $M_{BH} \gtrsim 10^8 M_{\odot}$. This is not necessarily wrong. \citet{Graham2013} suggest that there is a break in black hole scaling relations because the growth of $M_{BH}$ is dominated by different processes at different masses. As discussed previously, a break in the $M_{BH}-\sigma$ relation is also suggested by \citet{Martin2018}. Recent theoretical studies by \citet{Bustamante2019} also suggest that a transition near $M_{BH} \gtrsim 10^8 M_{\odot}$ should naturally emerge as a result of black hole spin evolution. \citet{Bustamante2019} also identify the same ``bottleneck'' effect at this specific BH mass scale which results in an overabundance of SMBHs at around $10^8 M_{\odot}$ at $z=0$.

Our analysis suggests that if the transition between quasar mode and low state mode feedback is indeed dependent on $M_{BH}$ as is implemented in TNG100, we may be able to find features associated with the critical $M_{BH}$ in the observed data. More specifically, if low state mode feedback is indeed more effective than quasar mode in nature, then black hole scaling relations should show similar over-density near the transition mass as is shown here.

We note that when comparing simulations with observations, quantities directly measured from the simulation data can be different from those inferred from observations. Thus when possible, we always use the quantities measured from synthetic observations of the simulation data (e.g., $\sigma$ and Sersic indices, see Table~\ref{table:data} for summary). In addition, different observations have different selection functions, which can bias the results. For example, the best fit $M_{BH}-\sigma$ relations in \citet{Kormendy2013, McConnell2013} and \citet{Martin2018Nature} have different slopes and/or normalizations from each other (Figure~\ref{fig:M_sigma}). Moreover, the observed sample may not have a well-defined selection function. For example, the data in \citet{Terrazas2016} and \citet{Graham2007} are not from single surveys, and thus the biases are hard to measure. For those reasons, comparisons between simulations and observations here are mostly done at a qualitative level. 

On the other hand, the simulation data are processed in the same way for Illustris and TNG100. As we have shown, black hole - host galaxy scaling relations can be quite different between the two simulations, and much of the difference can be attributed to how black holes physics is modeled. In Section~\ref{sec:results2}, we further explore how sub-grid models affect the complex interplay of black hole growth, star formation history and galaxy properties.

\section{SMBH Over-massiveness and quiescence}\label{sec:results2}
The $M_{BH}-\sigma$ relation is quite tight in both observations and simulations as is shown in Figure~\ref{fig:M_sigma}. However, there is still some amount of scatter \citep{Tremaine2002, Kayhan2009, McConnell2013} \footnote{It has been argued that the scatter in the $M_{BH}-\sigma$ relation is no larger than measurement error alone \citep{Ferrarese2000}.}. \citet{Martin2018Nature} measure the star formation histories of nearby massive galaxies ($\sigma >100 \rm km/s$) from their integrated optical spectra, and find that the host galaxies of over-massive SMBHs (defined as ones above the average $M_{BH}-\sigma$ relation) have formed earlier, and have had suppressed star formation for longer. In this section, we examine the relation between SMBH over-massiveness and quiescence of host galaxies in Illustris (Section~\ref{sec:results2a}) and TNG100 (Section~\ref{sec:results2b}).

\subsection{SMBH Over-massiveness in Illustris}\label{sec:results2a}

\begin{figure*}
\begin{center}
\includegraphics[scale=0.28,trim=0cm 0cm 0cm 0cm, clip=true]{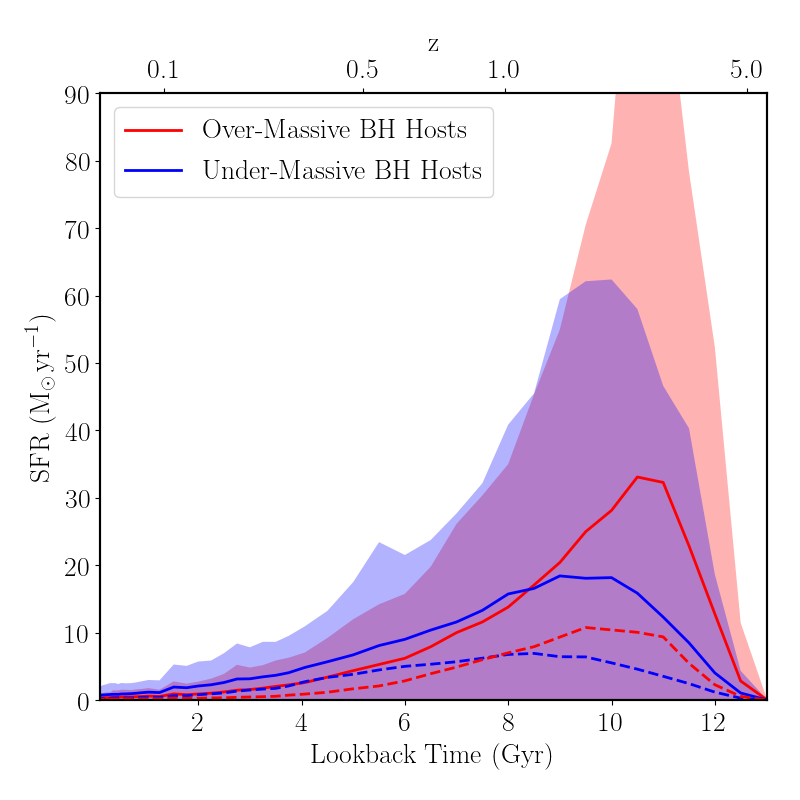} 
\includegraphics[scale=0.28,trim=0cm 0cm 0cm 0cm, clip=true]{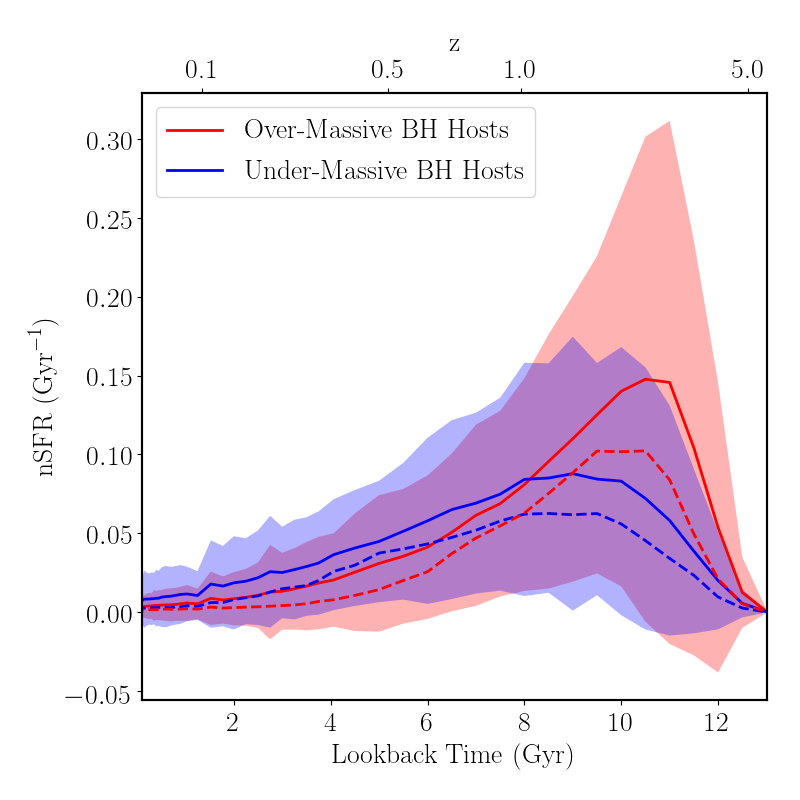} 
\includegraphics[scale=0.27,trim=0cm 0cm 0cm 0cm, clip=true]{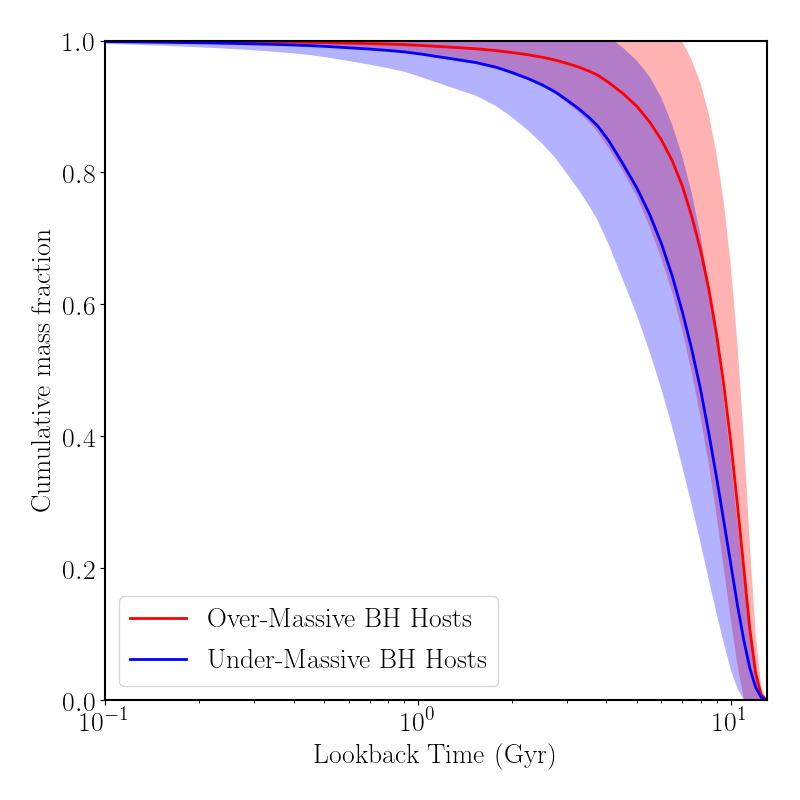}
\caption{Star formation history of massive galaxies ($\sigma >100 \rm km/s$) in Illustris. Top panel shows star formation rate as a function of lookback time. Red shows the galaxies with over-massive SMBHs and blue shows galaxies with under-massive SMBHs as defined in Section~\ref{sec:results2a}. Within each lookback time bin, we fit a normal distribution to the SFR distribution of each population, and the mean is shown as solid lines. The shaded regions are the 1 $\sigma$ range. Dashed lines show the median values for the two populations. Bottom panel shows the cumulative mass fraction as a function of lookback time, color-coded in the same way as the top panel. Similar to the observed massive galaxies in \citet{Martin2018Nature}, the hosts of over-massive SMBHs have formed earlier and have had suppressed star formation for longer in Illustris. 
\label{fig:SFH_Illustris}}
\end{center}
\end{figure*}

\begin{figure*}
\begin{center}
\includegraphics[scale=0.36,trim=0cm 0cm 0cm 0cm, clip=true]{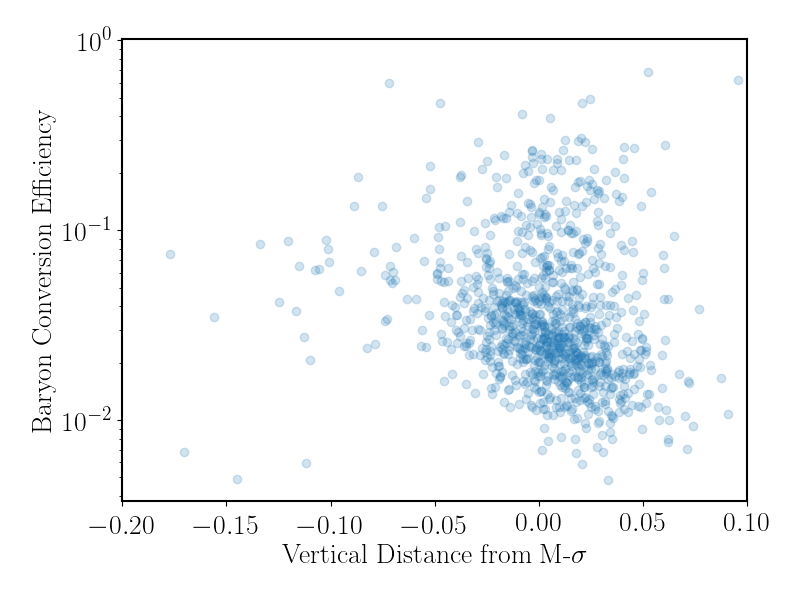}
\includegraphics[scale=0.4,trim=1cm 1cm 1cm 0cm, clip=true]{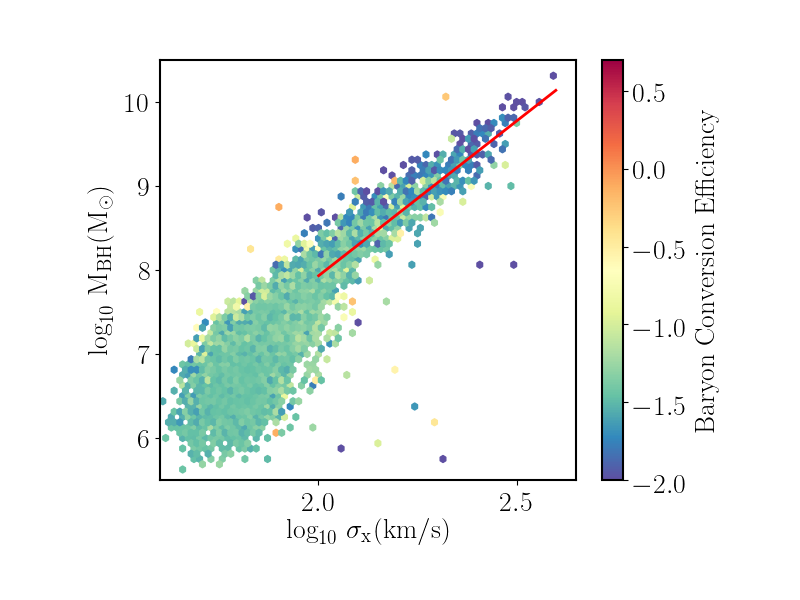}\\
\includegraphics[scale=0.36,trim=0cm 0cm 0cm 0cm, clip=true]{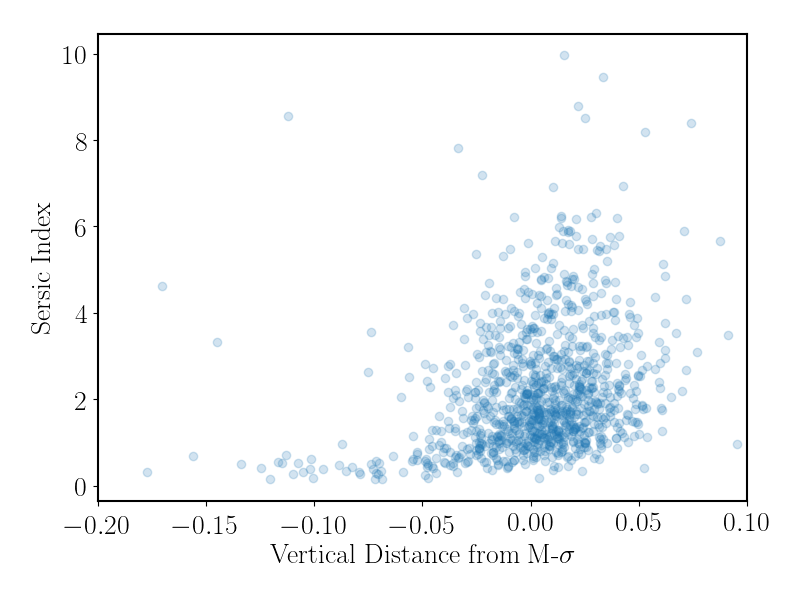}
\includegraphics[scale=0.4,trim=1cm 1cm 1cm 0cm, clip=true]{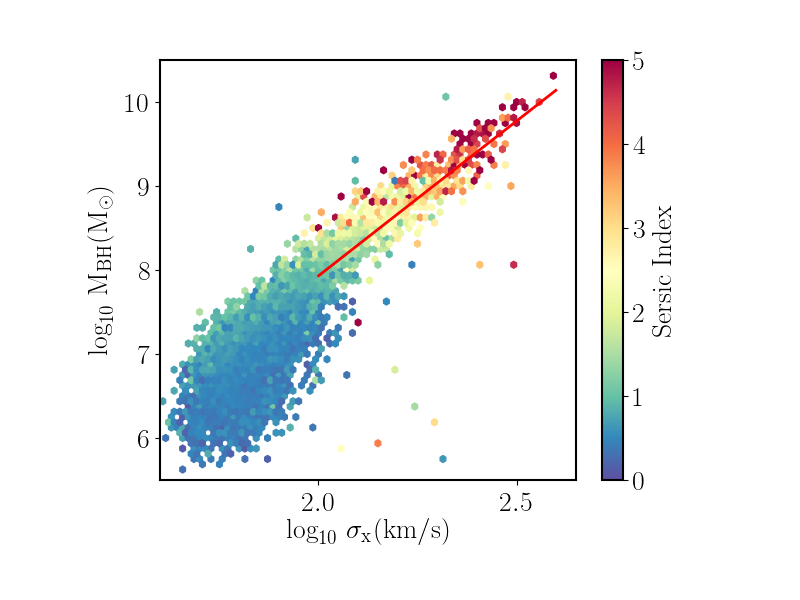} 
\caption{In Illustris, the over-massiveness of SMBHs (vertical distance from the average $M_{BH}-\sigma$ relation) correlates with the Sersic indices of the host galaxies, and anti-correlates with the baryon conversion efficiency. The left panels are showing only galaxies with $\sigma >100 \rm km/s$ in Illustris.
\label{fig:Illustris_weighted}}
\end{center}
\end{figure*}

\begin{figure}
\begin{center}
\includegraphics[scale=0.45,trim=0cm 0cm 0cm 6cm, clip=true]{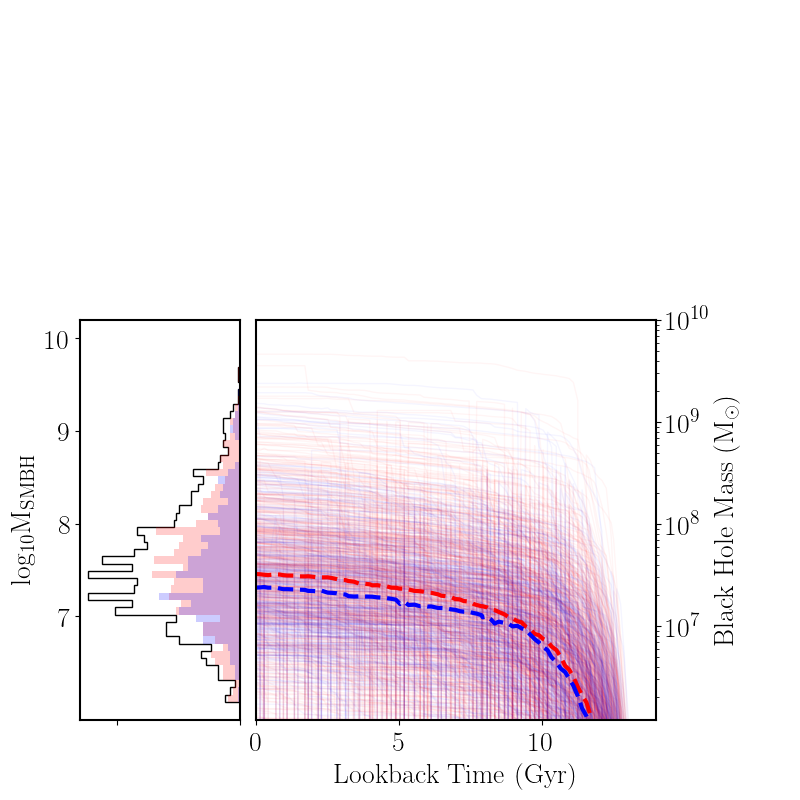} 
\caption{Main panel: the evolution of $M_{BH}$ (bottom) of galaxies with $\sigma >100 \rm km/s$ in Illustris. Red shows the galaxies with over-massive SMBHs and blue shows galaxies with under-massive SMBHs. Faint lines show the trajectory of individual galaxies. Thick dashed lines connect the median values of the two populations at each lookback time. The dips (downward trend followed by a sudden increase) represent a mis-identification of the main progenitor, which can occur during mergers \citep{Rodriguez-Gomez2015}. Small panel on the left shows the distribution of $M_{BH}$ at $z=0$. Again, red shows the galaxies with over-massive SMBHs and blue shows galaxies with under-massive SMBHs. The black line shows the mass distribution of all the SMBHs in galaxies with $\sigma >100 \rm km/s$ in Illustris. 
\label{fig:Illustris_history}}
\end{center}
\end{figure}

\begin{figure*}
\begin{center}
\includegraphics[scale=0.28,trim=0cm 0cm 0cm 0cm, clip=true]{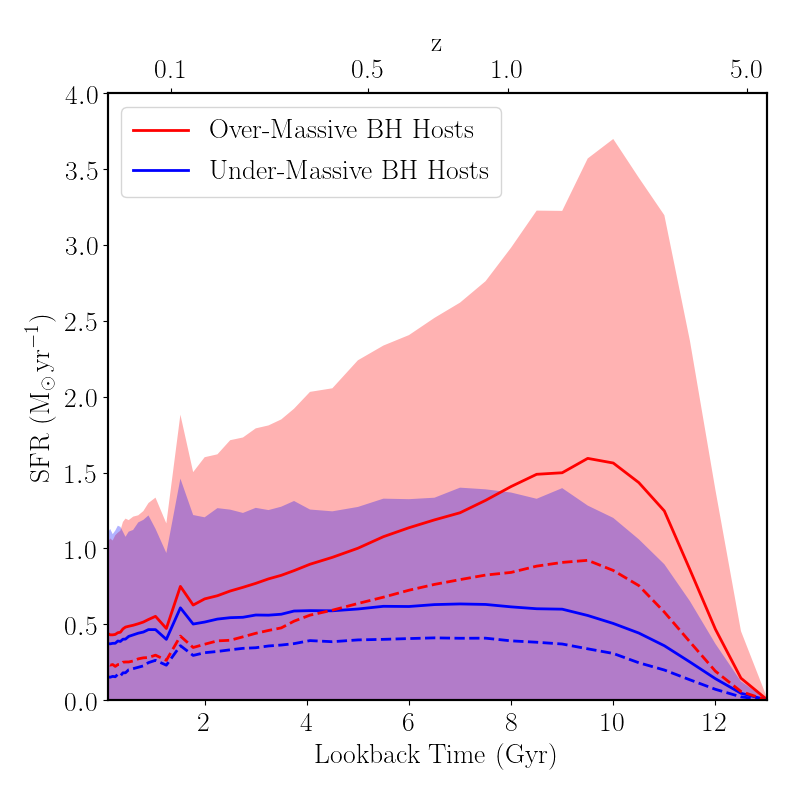}
\includegraphics[scale=0.28,trim=0cm 0cm 0cm 0cm, clip=true]{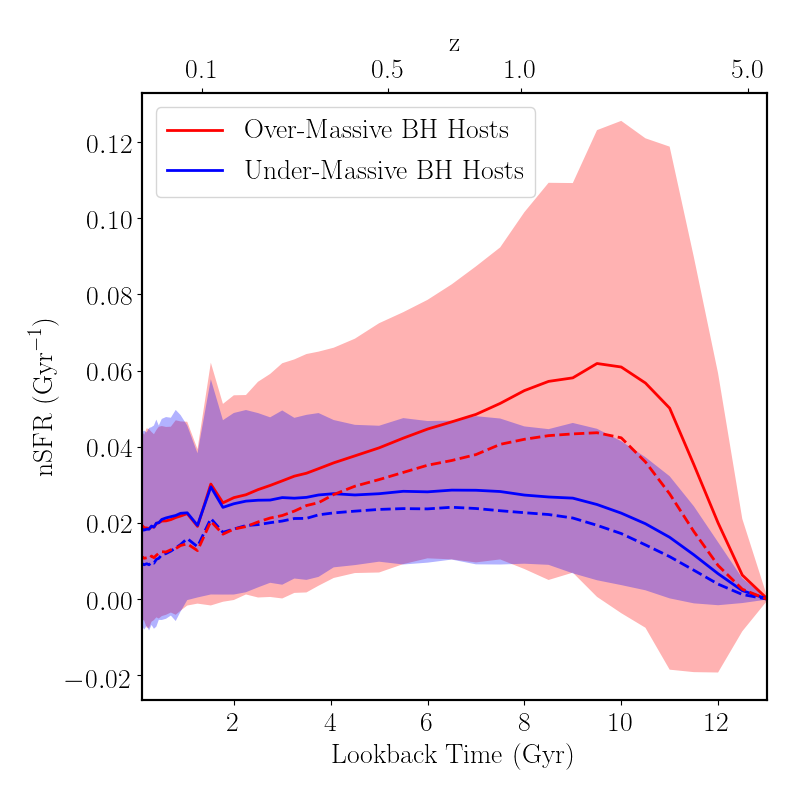} 
\includegraphics[scale=0.27,trim=0cm 0cm 0cm 0cm, clip=true]{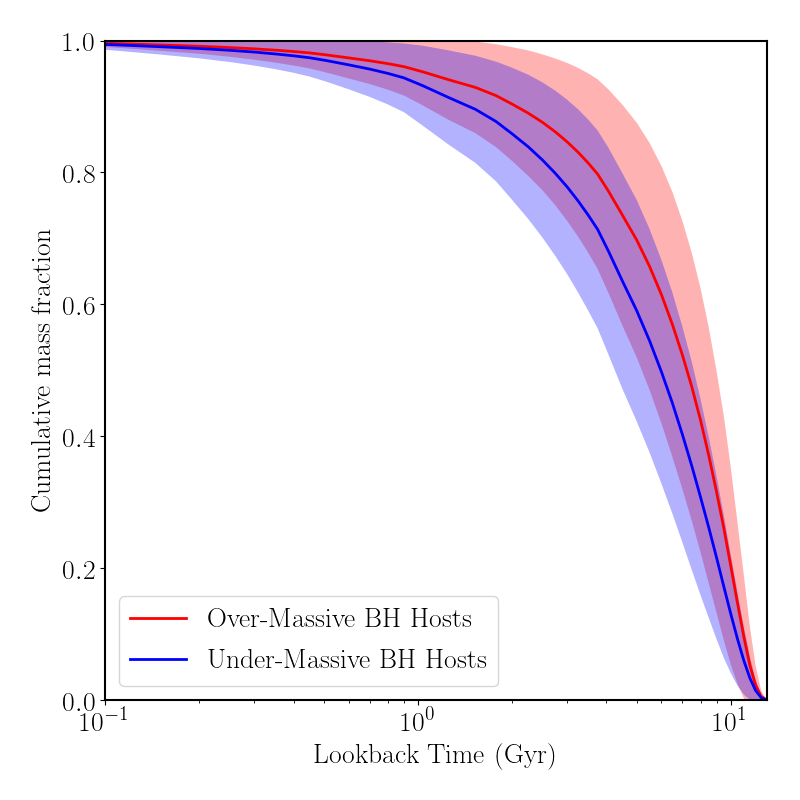}
\caption{Star formation history in low-mass galaxies with $\sigma <100 \rm km/s$ in Illustris. Colors and lines have the same meanings as Figure~\ref{fig:SFH_Illustris}. Similar to the massive galaxies in Illustris, for low-mass galaxies, the hosts of over-massive SMBHs have also formed earlier and have had suppressed star formation for longer than the hosts of under-massive SMBHs. This is different from \citet{Martin2018} which find no such correlation for low-mass galaxies in the observed sample. 
\label{fig:SFH_low}}
\end{center}
\end{figure*}

\subsubsection{SMBH Over-massiveness in massive galaxies in Illustris}
As Figure~\ref{fig:M_sigma} shows, there is an offset between simulated and observed $M_{BH}-\sigma$ relations. Thus, instead of using the observed $M_{BH}-\sigma$ relation, we first perform a linear fit to the simulated galaxies with $\sigma >100 \rm km/s$ and use that as the average $M_{BH}-\sigma$ relation to define over-massive and under-massive black holes in the two simulations separately. The best fitting $M_{BH}-\sigma$ relation for galaxies with $\sigma >100 \rm km/s$ in Illustris is $\rm log_{10}M_{BH}(\rm M_\odot) = 3.7 log_{10}\sigma(km/s) + 0.56 $. 

As discussed in Section~\ref{sec:results1}, both Illustris and TNG100 simulations likely under-predict $\sigma$ (by order of $\sim 0.1$ dex). \citet{Martin2018Nature} and \citet{Martin2018} use $\sigma=100 \rm km/s$ as the dividing line between massive and low-mass galaxies in the observed sample. When we use the same $\sigma=100 \rm km/s$ as the dividing line, we are effectively using a slightly higher $\sigma$, corresponding to a higher $M_{BH}$. We have tested using $\rm log_{10}\sigma (km/s)=1.9$ as the cut, and found that the results remain qualitatively the same for for both massive and low-mass galaxies for both Illustris and TNG100. Thus for simplicity and consistency, we use $\sigma=100 \rm km/s$ but caution that this selection criterion is slightly different from \citet{Martin2018Nature} due to the difference in $\sigma$ between simulations and observations. 

Figure~\ref{fig:SFH_Illustris} shows the comparison between the star formation histories of the massive galaxies hosting over-massive SMBHs (red) and under-massive SMBHs (blue) in Illustris. The left and middle panels show the absolute and normalized star formation rate (nSFR) as a function of lookback time and redshift. The nSFR is computed as star formation rate divided by the galaxy's current-day stellar mass, as is used in \citet{Martin2018Nature}. This is different from specific star formation rate (sSFR) as sSFR at any given time is calculated as the star formation rate divided by the galaxy's stellar mass at that time. The star formation rate of individual galaxies in the simulations is computed based on the formation histories of all the star particles that reside within the galaxy at $z=0$. This is different from the in-situ SFR of the galaxy progenitors as our SFR also includes the star particles that formed ex-situ, but were accreted later. We choose this way of computing SFR because this is the closest to what can be obtained from the spectra of the observed galaxies. 

The solid lines in Figure~\ref{fig:SFH_Illustris} are the mean SFR within each lookback time bin for the two populations, and the shaded area brackets the 1-$\sigma$ deviation. The dashed lines are the median values within each bin. Whether we use the mean SFR or median SFR, the host galaxies of over-massive SMBHs have a higher SFR and nSFR at higher redshift, and lower nSFR at lower redshift. The transition happens at $z\gtrsim 1$. The right panel shows the cumulative stellar mass as a function of lookback time. Galaxies hosting over-massive SMBHs have formed their stellar mass earlier than the ones hosting under-massive SMBHs. This is in qualitative agreement with the findings in \citet{Martin2018Nature} for the observed massive galaxies. Even the time that the nSFRs of the two populations cross is similar. In \citet{Martin2018Nature}, it happens about 10 Gyr ago, and in Illustris, about 8 Gyr ago. 

\citet{Martin2018Nature} has a selection bias due to line contamination. Galaxies with high current-day SFR tend to be excluded from the sample, and thus the sample is heavily biased toward quiescent galaxies. 
In order to test the effect of this selection bias, we have experimented with only selecting Illustris galaxies with sSFR $< 10^{-11} yr^{-1}$ and sSFR $< 10^{-12} yr^{-1}$. Out of our full Illustris sample of 6808 galaxies, these two criteria select 894 and 588 galaxies, respectively. Both subsamples show the same trend as the full sample. This confirms that the result is not sensitive to the selection bias related to SFR. For completeness, we use the full sample without a sSFR cut throughout the paper. \citet{Martin2018Nature} also use a fixed aperture of $0.5R_e$, where $R_e$ is half-light radius of a galaxy. We have adopted the same $0.5 R_e$ aperture, but our $R_e$ is the half-mass radius, which is not exactly the same as the half-light radius. We have experimented with $R_e$ and $2R_e$, and found the trend to be the same. This suggests that at least for simulated massive galaxies, the results are not sensitive to the exact aperture.

One interpretation of the correlation between SMBH over-massiveness and quiescence as proposed in \citet{Martin2018Nature} is that quenching happens earlier and more efficiently in galaxies hosting over-massive SMBHs. We use the vertical distance to the average $M_{BH}-\sigma$ as a measurement of over-massiveness of SMBHs, and define the baryon conversion efficiency as the total stellar mass divided by the total halo mass. We find a statistically significant negative correlation between the two quantities with a Pearson's correlation coefficient of $r=-0.12$ and a p-value of $log_{10}p\sim-4$ (top left panel of Figure~\ref{fig:Illustris_weighted}). This is also seen in the top right panel of Figure~\ref{fig:Illustris_weighted}, which shows the $M_{BH}-\sigma$ relation weighted by the baryon conversion efficiency. The galaxies that lie above the average $M_{BH}-\sigma$ tend to be less efficient in converting baryons into stars. \citet{Snyder2015} find that in Illustris, SMBHs that are over-massive with respect to the mean $M_{BH}$ - halo mass relation correlate with smaller host stellar mass. The interpretation is that higher SMBH mass implies more total feedback energy and thus a greater reduction in $M_*$. This is in concordance with our findings here.

Sersic indices of galaxies measure the concentration of stars in the host galaxy, and are correlated with the formation time of galaxies in that older galaxies tend to have higher Sersic indices. Given that the hosts of over-massive SMBHs have formed their stars earlier, one would expect them to have higher Sersic indices. The bottom two panels of Figure~\ref{fig:Illustris_weighted} show that over-massiveness of SMBHs is indeed positively correlated with the Sersic index of the host galaxy, with a Pearson's correlation coefficient of 0.22 and $log_{10}p\sim-12$. This is not surprising given that $M_{BH}$ is correlated with Sersic indices as is shown in Figure~\ref{fig:M_sigma}. However, we do note that the $M_{BH}-n$ relation does not necessarily mean that the over-massiveness on $M_{BH}-\sigma$ should be correlated with Sersic $n$.

When we trace the history of the main progenitors of the Illustris galaxies using the SubLink merger tree \citep{Rodriguez-Gomez2015}, we find that indeed the host galaxies of over-massive SMBHs have formed earlier and on average always have a slightly higher total halo mass. Figure~\ref{fig:Illustris_history} shows that the current-day over-massive SMBHs have also formed earlier and have been more massive throughout history. There is overlap between the hosts of over-massive SMBHs and the hosts of under-massive SMBHs, but the median values (thick dashed lines in Figure~\ref{fig:Illustris_history}) are well separated.

\subsubsection{SMBH Over-massiveness in low-mass galaxies in Illustris}

Although at $\sigma >100 \rm km/s$, the host galaxies of over-massive SMBHs and under-massive SMBHs have different star formation histories, \citet{Martin2018} find that for smaller galaxies with $\sigma <100 \rm km/s$, this is not the case. The star formation history of smaller galaxies appears uncorrelated with the over-massiveness of their SMBHs. Their interpretation is that AGN feedback is not important in small galaxies with $\sigma <100 \rm km/s$. We fit a linear $M_{BH}-\sigma$ relation to the galaxies with $\sigma <100 \rm km/s$ in Illustris \footnote{As discussed in Section~\ref{sec:results1}, the slope of $M_{BH}-\sigma$ is shallower for low-mass galaxies in \citet{Martin2018} but steeper in Illustris.} and define over-massive and under-massive SMBHs similarly to what is done for massive galaxies. We find that the over-massiveness of SMBHs in low-mass galaxies correlates with the star formation history in similar ways as the massive galaxies in Illustris. As Figure~\ref{fig:SFH_low} shows, the host galaxies of over-massive SMBHs have formed earlier, and their SFR is higher at higher redshift, but falls under the hosts of under-massive SMBHs. The nSFR of the two populations are comparable below $z\sim 2$. For low-mass galaxies, \citet{Martin2018} use a fixed SDSS aperture, which is about 4.5 kpc. We have adopted the same aperture for our low-mass galaxies here, which is roughly their average $R_e$. Since simulated galaxies in Illustris tend to have larger sizes than the observed galaxies at given stellar mass \citep{Bottrell2017}, using an aperture with a fixed physical size may create a selection bias. We have experimented with $0.5 R_e$ and $2R_e$, and found that the exact choice of aperture mainly affects the absolute level of SFR but does not change the basic trends. 

Similar to the massive galaxies in Illustris, the vertical distance to the $M_{BH}-\sigma$ relation for low-mass galaxies is also positively correlated with the Sersic index. This can be seen from the right panels of Figure~\ref{fig:Illustris_weighted}. However, the correlation between SMBH over-massiveness and baryon conversion efficiency for low-mass galaxies is different. Our statistical analysis suggests that they are positively correlated with each other with $r=0.14$ and $log_{10} p \sim -25$. 

Simulated low-mass galaxies in Illustris have a stronger coupling between star formation history and black hole mass than what is reported in \citet{Martin2018} for the observed low-mass galaxies.
There are a few possible explanations for this disagreement. Perhaps AGN feedback in smaller galaxies should be modeled differently than in more massive systems. It is also possible that the stochasticity of SMBH feeding and feedback, which is poorly captured by the simulation, has a bigger impact in low-mass galaxies due to their shallower gravitational potential. 
The disagreement may also be related to uncertainties and biases in the observations. For the observed galaxies, the uncertainties are larger for the estimates of both $M_{BH}$ and $\sigma$ at the low-mass end. The \citet{Martin2018} sample is based on single epoch BH mass determination of active galaxies. It is also possible that there are unknown biases in the analysis of the star formation history for low-mass galaxies. In addition, the low-mass galaxies analyzed in the simulations are effectively selected based on their stellar masses at the low-mass end, while the selection bias in the observed sample is more complicated \citep{Martin2018, Shankar2019}. 

Recently, \citet{Dickey2019} conducted a spectroscopic study of a sample of isolated low-mass galaxies, and concluded that AGN feedback is likely responsible for quenching these galaxies. \citet{Penny2018} and \citet{Manzano2019} find AGN-driven outflows from a sample of dwarf galaxies, and evidence of suppressed star formation. These findings suggest that AGN feedback may still play a role in at least some low-mass galaxies, and are consistent with what we find here based on the Illustris simulation. Note that in general, quenching in Illustris is found to be not efficient enough compared with the observations, whereas the level of star formation in TNG100 is in better agreement with the observations \citep{Bluck2016, Nelson2018, Donnari2019}. Our discussions here are mainly focused on the relative effect of feedback between different galaxies within the simulation, not the absolute level of quenching.

\subsection{SMBH Over-massiveness in TNG100}\label{sec:results2b}
\begin{figure*}
\begin{center}
\includegraphics[scale=0.28,trim=0cm 0cm 0cm 0cm, clip=true]{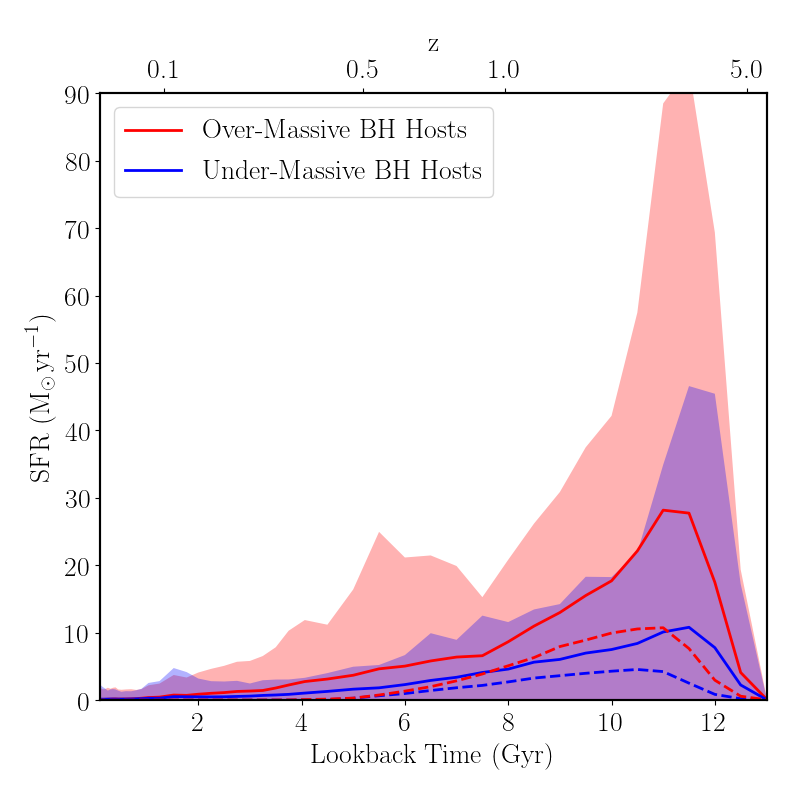} 
\includegraphics[scale=0.28,trim=0cm 0cm 0cm 0cm, clip=true]{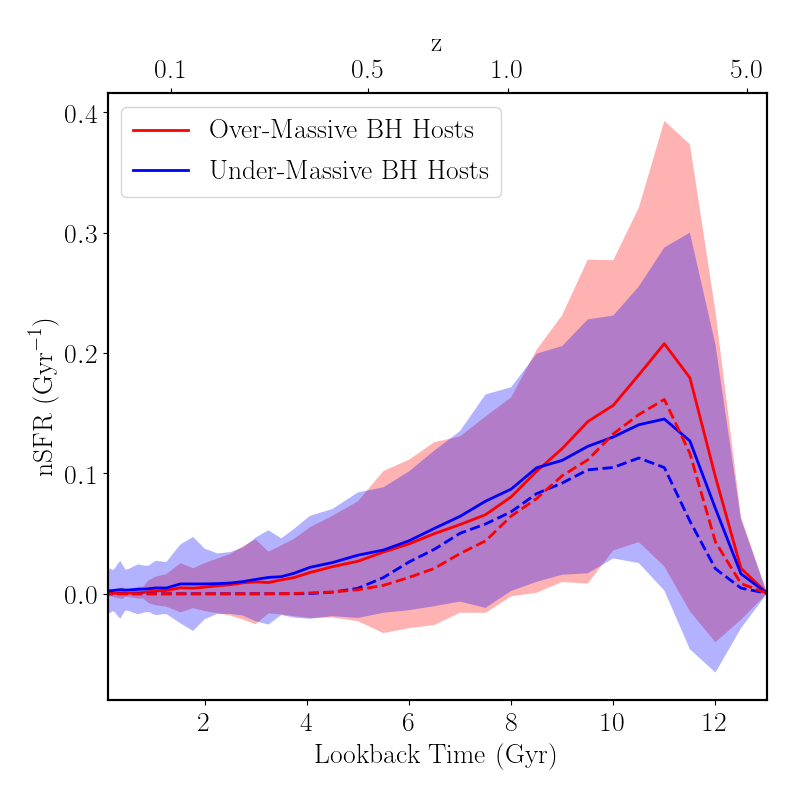} 
\includegraphics[scale=0.27,trim=0cm 0cm 0cm 0cm, clip=true]{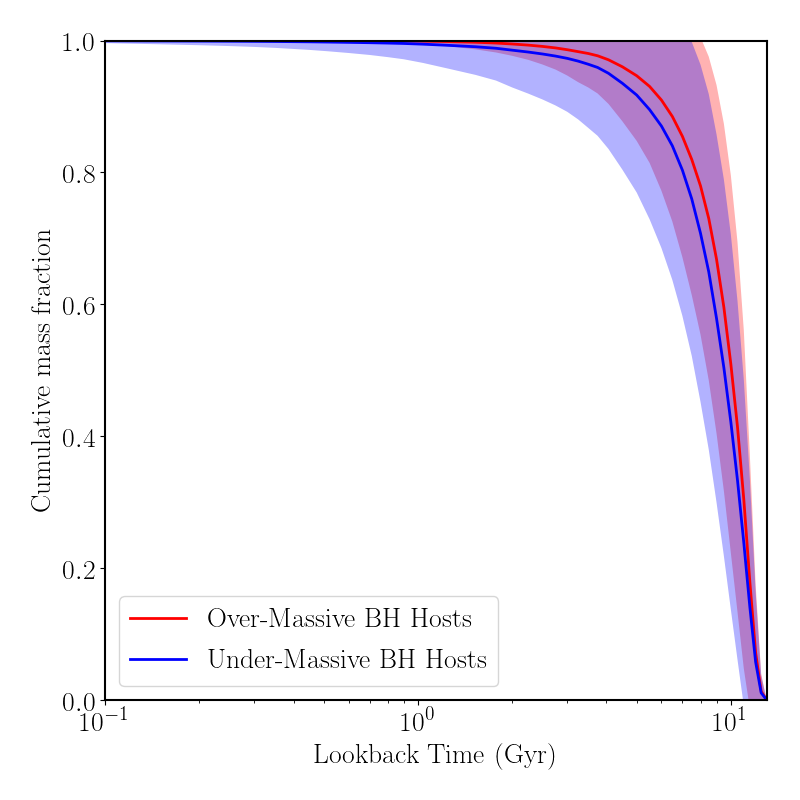}
\caption{Star formation history of massive galaxies ($\sigma >100 \rm km/s$) in TNG100. Colors and lines have the same meanings as Figure~\ref{fig:SFH_Illustris}. The correlation between SMBH over-massiveness and the host star formation history is different in TNG100 than in Illustris and the observations in \citet{Martin2018}. On average, the hosts of over-massive SMBHs in TNG100 have higher SFR than the hosts of under-massive SMBHs throughout history, but the two populations have similar nSFR and cumulative star formation history. 
\label{fig:SFH_TNG}}
\end{center}
\end{figure*}

\begin{figure}
\begin{center}
\includegraphics[scale=0.45,trim=0cm 0cm 0cm 6cm, clip=true]{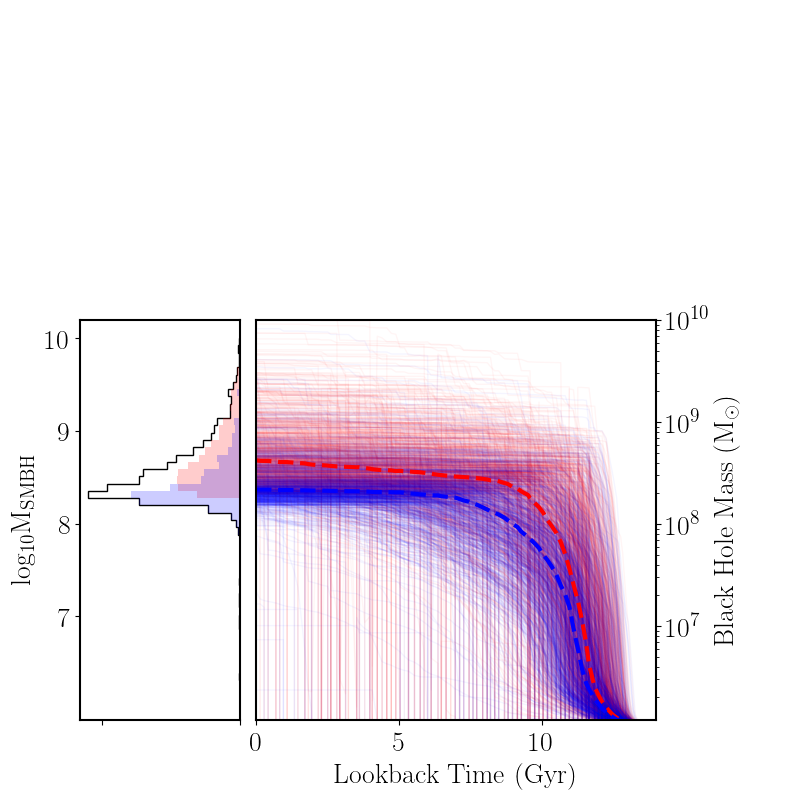}
\caption{Main panel: the growth history of over-massive SMBHs (red) and under-massive SMBHs (blue) in TNG100. Left panel shows the distribution of $M_{BH}$ at $z=0$. Thick dashed lines show the median values of the two populations. This is to be compared with Figure~\ref{fig:Illustris_history} for Illustris. Faint lines show the trajectory of individual galaxies, and are slightly fainter than in Figure~\ref{fig:Illustris_history} for the clarity of the figure. 
\label{fig:TNG_history}}
\end{center}
\end{figure}

\begin{figure}
\begin{center}
\includegraphics[scale=0.4,trim=0cm 1cm 2cm 0cm, clip=true]{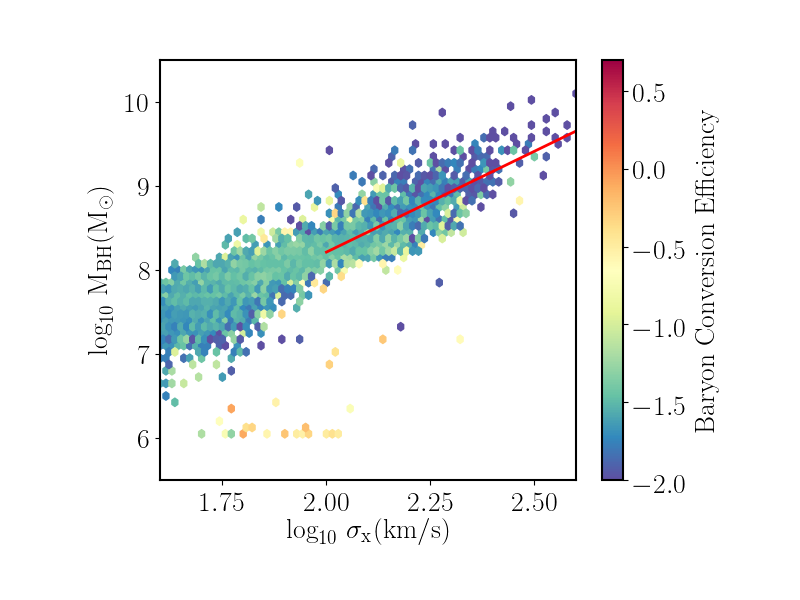}
\includegraphics[scale=0.4,trim=0cm 1cm 2cm 0cm, clip=true]{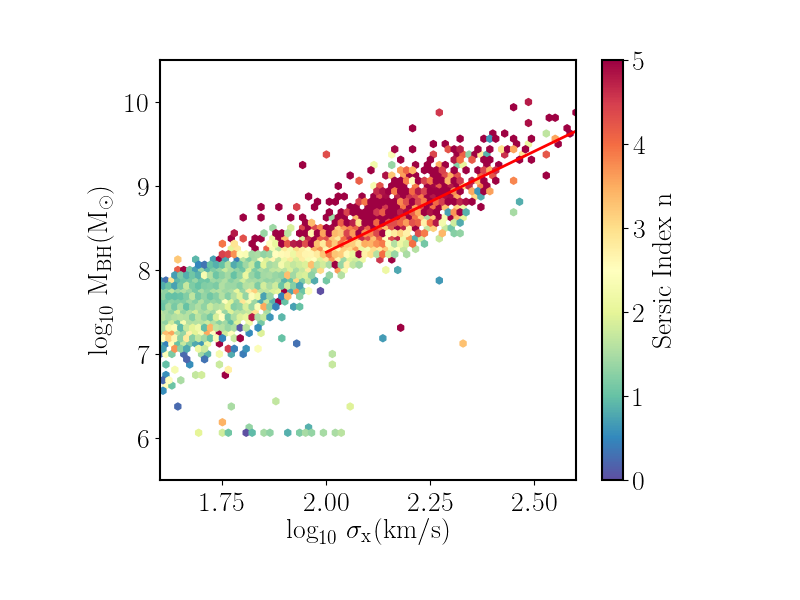} 

\caption{$M_{BH}-\sigma$ relation weighted by baryon conversion efficiency (top) and Sersic indices (bottom) in TNG100.
\label{fig:TNG_weighted}}
\end{center}
\end{figure}

\begin{figure*}
\begin{center}
\includegraphics[scale=0.28,trim=0cm 0cm 0cm 0cm, clip=true]{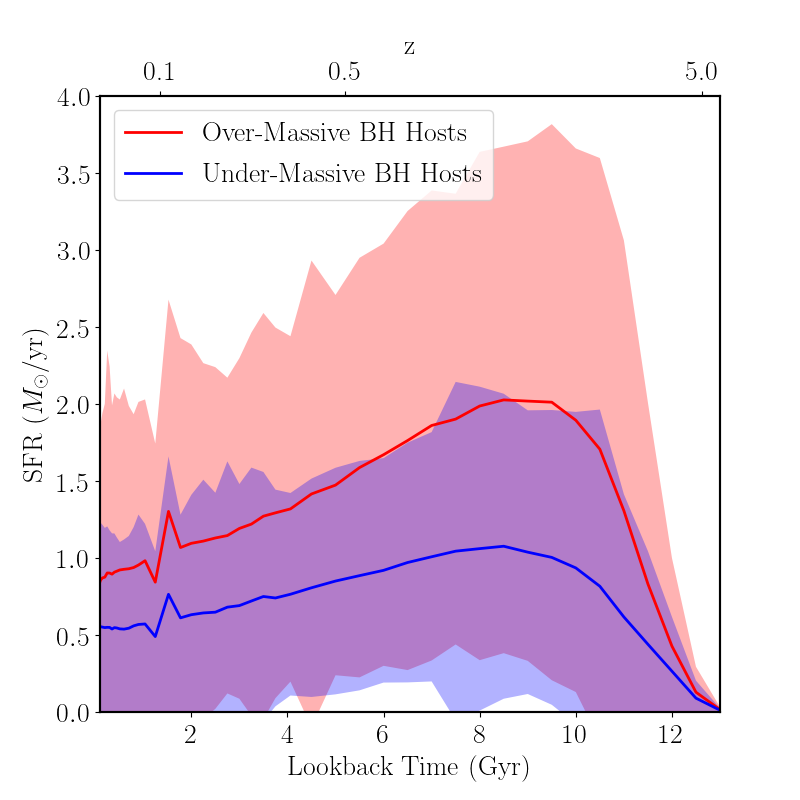} 
\includegraphics[scale=0.28,trim=0cm 0cm 0cm 0cm, clip=true]{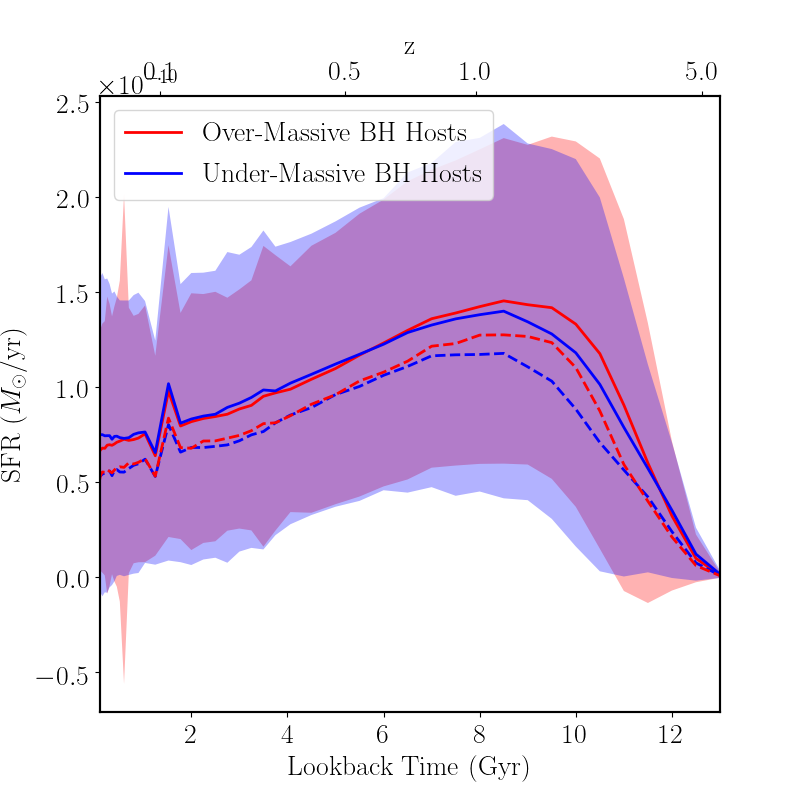} 
\includegraphics[scale=0.27,trim=0cm 0cm 0cm 0cm, clip=true]{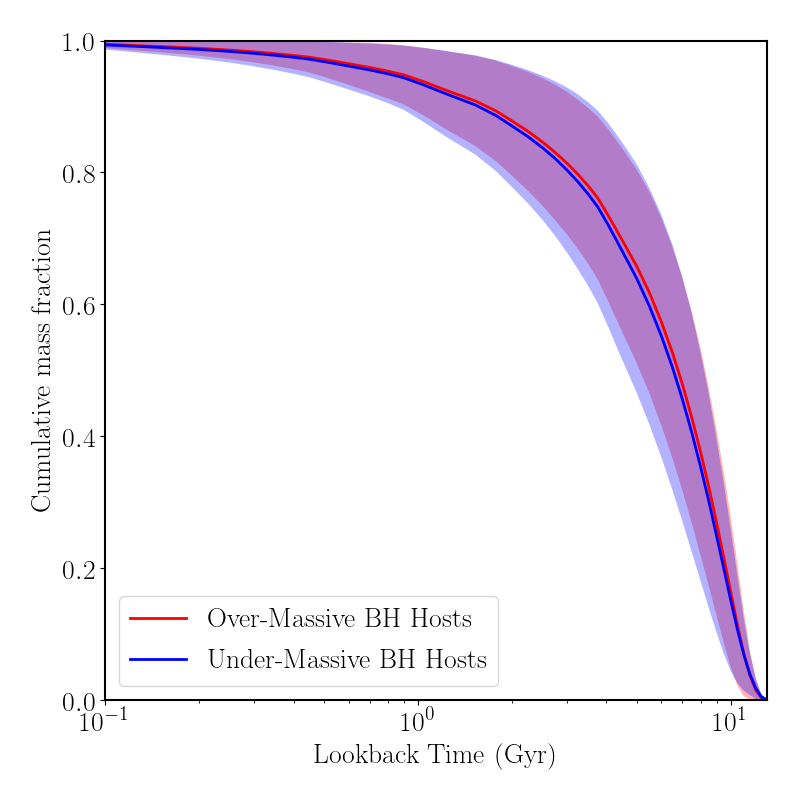}
\caption{Star formation history of low-mass galaxies ($\sigma <100 \rm km/s$) in TNG100. Colors and lines have the same meanings as Figure~\ref{fig:SFH_Illustris}. Similar to the massive galaxies in TNG100, for low-mass galaxies, the hosts of over-massive SMBHs have higher SFR than the hosts of under-massive SMBHs throughout history, and the normalized star formation histories of the two populations are very similar. 
\label{fig:SFH_TNG_low}}
\end{center}
\end{figure*}

We have performed the same analysis with the TNG100 data. The best linear fit $M_{BH}-\sigma$ relation for massive galaxies in TNG100 is $\rm log_{10}M_{BH}(\rm M_\odot) = 2.4log_{10}\sigma(km/s) + 3.4$. As discussed in Section~\ref{sec:results2a}, in both observations and Illustris simulation, for galaxies with $\sigma >100 \rm km/s$, the host galaxies of over-massive SMBHs have formed earlier and have been more quiescent at low redshift. This is not the case for TNG100. 

Figure~\ref{fig:SFH_TNG} shows that in TNG100, massive galaxies hosting over-massive SMBHs have always had a higher SFR than the hosts of under-massive black holes throughout their star formation history. They also have a slightly higher nSFR than the hosts of under-massive black holes at high redshift, but the nSFR of the two populations become similar at $z\lesssim1$. The formation time of the hosts of over-massive SMBHs, measured by the cumulative mass fraction, is slightly earlier but the difference is barely noticeable. We have also compared the history of sSFR, and the hosts of under-massive SMBHs have higher sSFR than the hosts of over-massive SMBHs at low z, but only by a very small amount. At $z=0$, the median sSFR of the hosts of over-massive SMBHs is $\sim 3\times 10^{-12}/yr$ and for the hosts of under-massive SMBHs, it is $\sim 4 \times 10^{-12}/yr$. For comparison, the median sSFRs of the hosts of over- and under-massive SMBHs in Illustris are $\sim 2 \times 10^{-12}/yr$ and $\sim 4 \times 10^{-11}/yr$, respectively. 

To test the effect of selection bias against star forming galaxies in \citet{Martin2018Nature}, we have carried out the same experiment on TNG100 as we did for Illustris: we select TNG100 galaxies with specific star formation rate (sSFR) $< 10^{-11} yr^{-1}$ and sSFR $< 10^{-12} yr^{-1}$, resulting in a subsample of 1780 and 1754 galaxies, respectively. Similar to the Illustris data, both TNG100 subsamples show the same trend as the full sample. Therefore we use the full sample without a sSFR cut throughout the paper for TNG100 as well. 

With a higher SFR throughout the history, the hosts of over-massive SMBHs should have a higher total stellar mass than the hosts of under-massive SMBHs. We have verified that the vertical distance to the average $M_{BH}-\sigma$ relation is strongly correlated with $M_*$ in TNG100. This is consistent with our finding in Section~\ref{sec:results1} that in TNG100, $M_{BH}$ and $M_*$ are very tightly correlated with each other. Thus larger $M_{BH}$ on the $M_{BH}-\sigma$ relation translates to larger $M_*$. 

Since a linear fit does not describe the $M_{BH}-\sigma$ relation in TNG100 very well (Figure~\ref{fig:M_sigma}), we have also tried to define the average $M_{BH}-\sigma$ relation using higher-order polynomials. We have also tried to divide the data into bins based on $\sigma$ and use the median $M_{BH}$ within $\sigma$ bins. All these experiments give the same results. We have also tried to choose a narrower range of $\sigma$. For example, we have selected only galaxies with $\rm 2.2<log_{10}\sigma (km/s) <2.4$ (the very high mass end avoiding the flattening part of the $M_{BH}-\sigma$ relation), and the correlation remains the same. When we select galaxies with $\rm 1.9<log_{10}\sigma (km/s) <2.1$ (the flattening part of $M_{BH}-\sigma$ around the threshold BH mass for the low state mode feedback), we find that the hosts of over-massive SMBHs do have a slightly lower average SFR than the hosts of under-massive SMBHs at $z=0$. However, this suppression of star formation happened only 1 Gyr ago, instead of several Gyr ago as is found in Illustris and the observed data.

Given the large uncertainties in recovering star formation history of galaxies observationally, there is a possibility that the observed correlation in \citet{Martin2018Nature} is due to hidden biases. It is uncanny that Illustris would reproduce similar trends though. This is even more intriguing considering that Illustris intermediate and high-mass galaxies have been demonstrated not to show the observed levels of  ``quiescence'', and they exhibit overall smaller quenched or red fractions than the TNG populations \citep{Vogelsberger2014, Nelson2018, Donnari2019}. More likely, TNG100 fails to recover the correlation between SMBH over-massiveness and quiescence.

One possible explanation for the discrepancy is the overly efficient feedback at $M_{BH}\gtrsim 10^8M_\odot$ discussed previously. The massive galaxies with $\sigma >100 \rm km/s$ have $M_{BH}$ above this critical mass and are in the very effective pure kinetic feedback mode described in Section~\ref{sec:simulations}. The effect of overly efficient feedback at the threshold mass can also be seen in Figure~\ref{fig:TNG_history}, which shows the evolution of SMBHs in galaxies with $\sigma >100 \rm km/s$ in TNG100. We see again the flattening of the history of individual galaxies at $M_{BH}\gtrsim 10^8M_\odot$. Note that the actual transition black hole mass is slightly higher at higher redshift (between lookback time of 5 and 10 Gyr, the over-dense ``strip'' turns upwards). This is simply because at higher redshift, BH accretion rates tend to be higher, and the transition to low state mode is harder (see Equation~\ref{eq:BH} and \citealt{Weinberger2018}). 

The galaxies hosting over-massive SMBHs and the ones hosting under-massive SMBHs are less separated in TNG100 than in Illustris, and the distance between the median values is also smaller (Figure~\ref{fig:Illustris_history}). The distribution of $M_{BH}$ in TNG100 is narrower and more peaked near the critical mass than that in Illustris. All of these are consistent with the shallower slope of the $M_{BH}-\sigma$ relation in TNG100 discussed earlier, and the horizontal features on BH scaling relations (Section~\ref{sec:results1}). 

Figure~\ref{fig:TNG_history} also suggests that the over-density around $M_{BH}\gtrsim 10^8M_\odot$ already exists more than 8 Gyr ago (corresponding to $z\sim 1$). In Illustris, this happens to be the time when hosts of under-massive SMBHs start to show higher average SFR than the hosts of over-massive SMBHs. The very effective kinetic feedback in TNG100 around the critical black hole mass not only limits the growth of the SMBHs themselves, but also very effectively suppresses star formation of the host galaxies. It is likely that some galaxies with under-massive SMBHs (but still with $M_{BH}\gtrsim 10^8M_\odot$) should have more late-time star formation, but end up producing too few young stars in TNG100. Another factor that contributes to the discrepancy between TNG100 and observations is that SMBHs also tend to be generally overly massive at $\rm log_{10}\sigma (km/s)<2.2$ (Section~\ref{sec:results1}). Thus we miss a population of galaxies with lower $M_{BH}$ and low-level star formation (but not fully quenched) at low redshift. This is consistent with what we find in \citep{Terrazas2019}. We emphasize again that our discussions are focused on a specific aspect of the observations that compare the relative differences in the star formation history of galaxies related to their black hole masses. In general, TNG100 produces quenched galaxies in good agreement with the observations in terms of their absolute level of star formation (or suppression of star formation), while Illustris galaxies tend to have a SFR that is too high \citep{Donnari2019}.

Although the correlation between SMBH over-massiveness and quiescence in TNG100 is different from Illustris, for massive galaxies, SMBH over-massiveness correlates with baryon conversion efficiency and Sersic indices in a similar way: galaxies with over-massive SMBHs tend to have a lower baryon conversion efficiency, and a larger Sersic index (Figure~\ref{fig:TNG_weighted}). However, this is not the case for low-mass galaxies in TNG100. As Figure~\ref{fig:TNG_weighted} shows, for galaxies with $\sigma <100 \rm km/s$, there is no obvious correlation between over-massiveness and Sersic $n$ or baryon conversion efficiency. A Pearson's correlation analysis also suggests no statistically significant correlations.

On the other hand, the star formation history of low-mass galaxies in TNG100 shows a very weak correlation with the over-massiveness of their SMBHs (Figure~\ref{fig:SFH_TNG_low}), very similar to TNG100 massive galaxies. In that regard, low-mass galaxies in TNG100 behave similarly to the observed low-mass galaxies in \citet{Martin2018}. However, we again note that the TNG100 $M_{BH}-\sigma$ relation differ from the observed relation in both its normalization and slope at the low-mass end. Low-mass galaxies in TNG100 also show a much tighter $M_{BH}-M_*$ relation than the observed galaxies (Figure~\ref{fig:M_sigma} and Section~\ref{sec:results1}).  

\subsection{Multi-dimensional SMBH scaling relations and the effects of AGN feedback}\label{sec:discussions}
Since the discovery of the $M_{BH}-\sigma$ relation, much attention has been paid to its scatter \citep{Gebhardt2000, Ferrarese2000, Kayhan2009}. It has been shown that the scatter in $M_{BH}-\sigma$ relation can be reduced by introducing a third parameter related to the property of the host galaxy \citep{Marconi2003, Barway2007, Graham2008}. In other words, similar to elliptical galaxies, there are ``fundamental planes'' for supermassive black holes as well. Our study also suggests that although the $M_{BH}-\sigma$ relation is rather tight, and is often considered ``fundamental'', it should not be viewed as a one-to-one correlation. Statistically, galaxies on the two sides of the mean relation have slightly different formation histories. In the Appendix, we show the $M_{BH}-\sigma-n$ and $M_{BH}-\sigma-eff$ (baryon conversion efficiency) planes for galaxies with $\sigma >100 \rm km/s$ in both Illustris and TNG100. In all cases, adding a third parameter reduces the residuals of the fit. In reality, the evolution of SMBHs correlates with many properties of the host galaxies, and the most fundamental black hole correlation is likely not a linear correlation or a plane, but rather, a multi-dimensional manifold. 

We also stress that the correlation between over-massiveness of SMBHs and quiescence seen in Illustris and the observed data \citep{Martin2018Nature} should not be simply attributed to ``effective'' AGN feedback (although it is certainly related). It is known that AGN feedback is not very effective in Illustris (in terms of shutting off late-time star formation), and is very effective for the massive galaxies in TNG100. Naively, one would expect to see a stronger correlation between over-massiveness of SMBHs and quiescence in TNG100 than Illustris, especially given that the effective kinetic feedback is linked to $M_{BH}$ in TNG100 by design. Instead, the observed trend is seen more strongly in Illustris which has less effective AGN feedback, but not TNG100 with more effective feedback. An important difference between AGN feedback in Illustris and TNG100 is that TNG100 feedback has more ``features'' (e.g., the switch between different feedback modes depends on $M_{SMBH}$). Some SMBH properties or correlations are more sensitive to the sheer strength of AGN feedback, while some, such as the one we examined here, are more sensitive to how the implementation of AGN feedback is related to the properties of the host galaxies. Our analysis shows that the strength of feedback does not guarantee an imprint in $M_{BH}-\sigma-$ vs. star formation history. BH growth, star formation history, stellar mass and $\sigma$ are related in complicated ways. As we have shown, the over-massiveness of SMBHs is related to many other properties of the host galaxies, including the Sersic indices and the baryon conversion efficiency. The galaxies hosting over-massive SMBHs in Illustris not only formed their stars earlier, but have also formed the SMBHs earlier, and have grown the dark matter halos earlier and bigger.

\section{Conclusions}\label{sec:conclusions}
We have studied the correlations between SMBH and the properties of their host galaxies in the Illustris and TNG100 simulations, including the $M_{BH}-\sigma$ relation, $M_{BH}-M_*$ relation, and the relation between $M_{BH}$ and Sersic index of the host galaxy. We have examined the correlation between SMBH over-massiveness (defined as the vertical distance to the average $M_{BH}-\sigma$ relation) and the star formation history of the host galaxies in Illustris and TNG100. We have contrasted their outcomes with those determined by selected observational datasets, albeit taken at face value and without correcting for observational selection biases. The main findings are summarized as follows:

\begin{enumerate}

\item In both Illustris and TNG100, $M_{BH}$ is correlated with $\sigma$, $M_*$, and the Sersic index, similar to the observations. The $M_{BH}-M_*$ relation is tighter in simulations than observations, especially in TNG100. In both Illustris and TNG100, the $M_{BH}-\sigma$ relation and $M_{BH}$-n relation show an offset from the observed correlations. This is due to the fact that the simulated galaxies host somewhat overmassive blackholes (by 0.2 dex for Illustris and 1.2 dex for TNG100 at $\rm log_{10}M_*(M_\odot)=10.2$), as well as may be too large (by up to 0.1 dex at $\rm M_*=10^{11} M_\odot$ ) and consequently have too small velocity dispersions (by up to ~0.05 dex at the same mass). We speculate that the offset at the low-mass end for TNG100 also has to do with the large SMBH seed mass combined with efficient growth (in-efficient feedback) in the TNG100 quasar mode.

\item All the black hole scaling relations in TNG100 show horizontal features at $M_{BH} \gtrsim 10^8 M_{\odot}$. This is the critical mass that helps define the transition between quasar-mode and low state mode feedback. If this transition has a $M_{BH}$ dependent Eddington-ratio transition instead of a fixed fraction of the Eddington ratio, we may be able to find similar features in the observed data. 

\item In Illustris, galaxies that host over-massive SMBHs (with $M_{BH}$ above the mean $M_{BH}-\sigma$ relation) on average have formed earlier (with higher star formation rate at high redshift), and have had lower star formation rates at low redshift compared with the hosts of under-massive SMBHs. This is in qualitative agreement with the findings in \citet{Martin2018Nature} for the observed massive galaxies with $\sigma >100 \rm km/s$. For galaxies with $\sigma <100 \rm km/s$, such a trend still exists in Illustris, but in the observations, star formation history does not show a clear correlation with the over-massiveness of SMBHs in low-mass galaxies \citep{Martin2018}.

\item In TNG100, galaxies that host over-massive SMBHs on average also tend to have formed earlier, but the trend is very weak. Galaxies with over-massive SMBHs in TNG100 always have a higher average star formation rate than the hosts of under-massive SMBHs throughout their star formation history. SMBH over-massiveness strongly correlates with $M_*$ of the host galaxies. The correlation between SMBH over-massiveness and the star formation history of the hosts is weaker in TNG100 possibly due to AGN feedback being too efficient at $M_{BH} \gtrsim 10^8 M_{\odot}$. This overly suppresses both the growth of the SMBHs themselves and star formation in these galaxies. This correlation reflects the interplay of BH growth, star formation history, and galaxy structure in a complex way, even in models where AGN feedback is responsible for quiescence. 

\item For all the galaxies we have studied with $\sigma >100 \rm km/s$ in Illustris and TNG100, over-massiveness of SMBHs correlates positively with the Sersic indices, and negatively with the baryon conversion efficiency. In both Illustris and TNG100, Sersic indices and baryon conversion efficiency can be used as a second parameter to reduce the amount of scatter in the $M_{BH}-\sigma$ relation for massive galaxies. The correlations are different for galaxies with $\sigma <100 \rm km/s$.

\end{enumerate}

The physical properties of a galaxy are determined by its full formation history, which is extremely complicated. We should interpret correlations with caution as they do not necessarily suggest causality or at least not direct or simple causality. One the other hand, the complex interplay of physical processes in models with effective low state mode AGN feedback can yield a diversity of observational signatures. These signatures are powerful and decisive probes of the implementation of diverse physical processes in hydro models. Therefore, observations of BH-host galaxy relations are powerful discriminants between models, and with sufficient care can be used to guide refinement of these models.

\section*{Acknowledgements}

We thank Dandan Xu for generously providing her catalogs. We thank Amy Reines and Alister Graham for sharing their observed data. We thank Eliot Quataert, Ena Choi, Kareem El-Badry, and Ignacio Martin-Navarro for helpful discussions. We acknowledge the technical support from the Scientific Computing Core of the Simons Foundation.

\vspace{0.2in}

%\bibliography{library}

\clearpage
\section*{Appendix}
\begin{figure*}
\begin{center}
\includegraphics[scale=0.5,trim=3cm 1cm 2cm 2cm, clip=true]{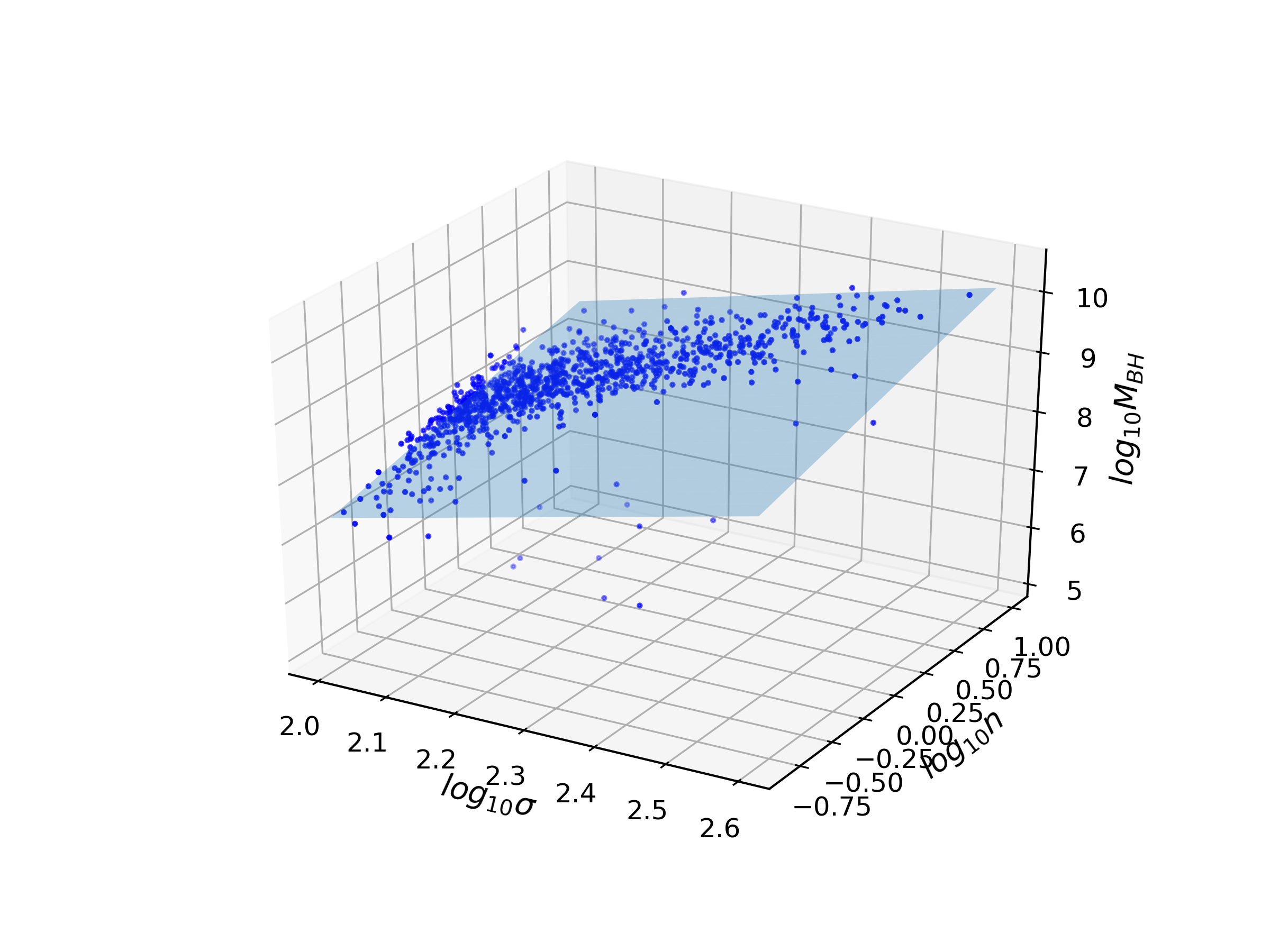}
\includegraphics[scale=0.5,trim=3cm 1cm 2cm 2cm, clip=true]{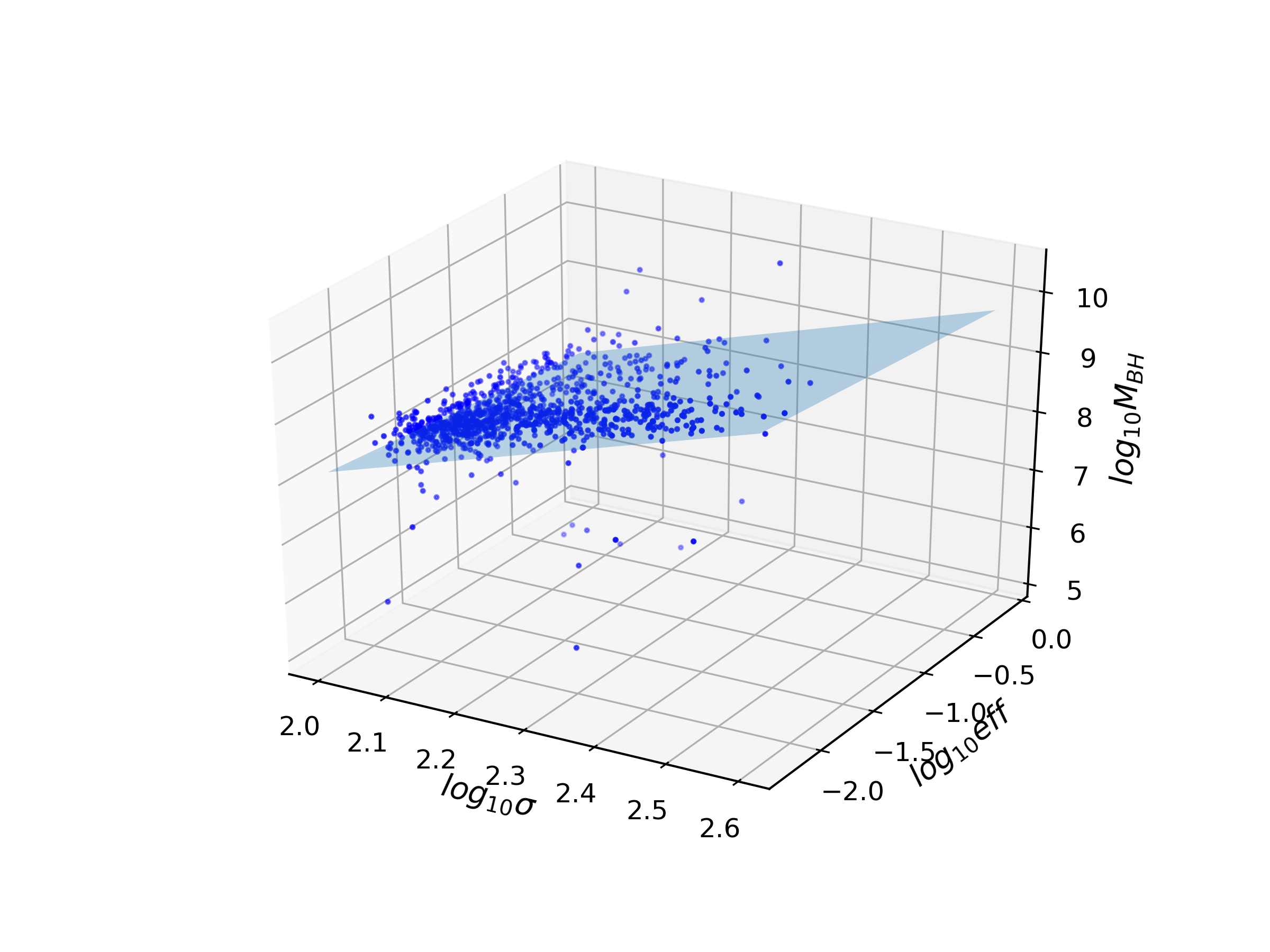}\\
\includegraphics[scale=0.5,trim=3cm 1cm 2cm 1cm, clip=true]{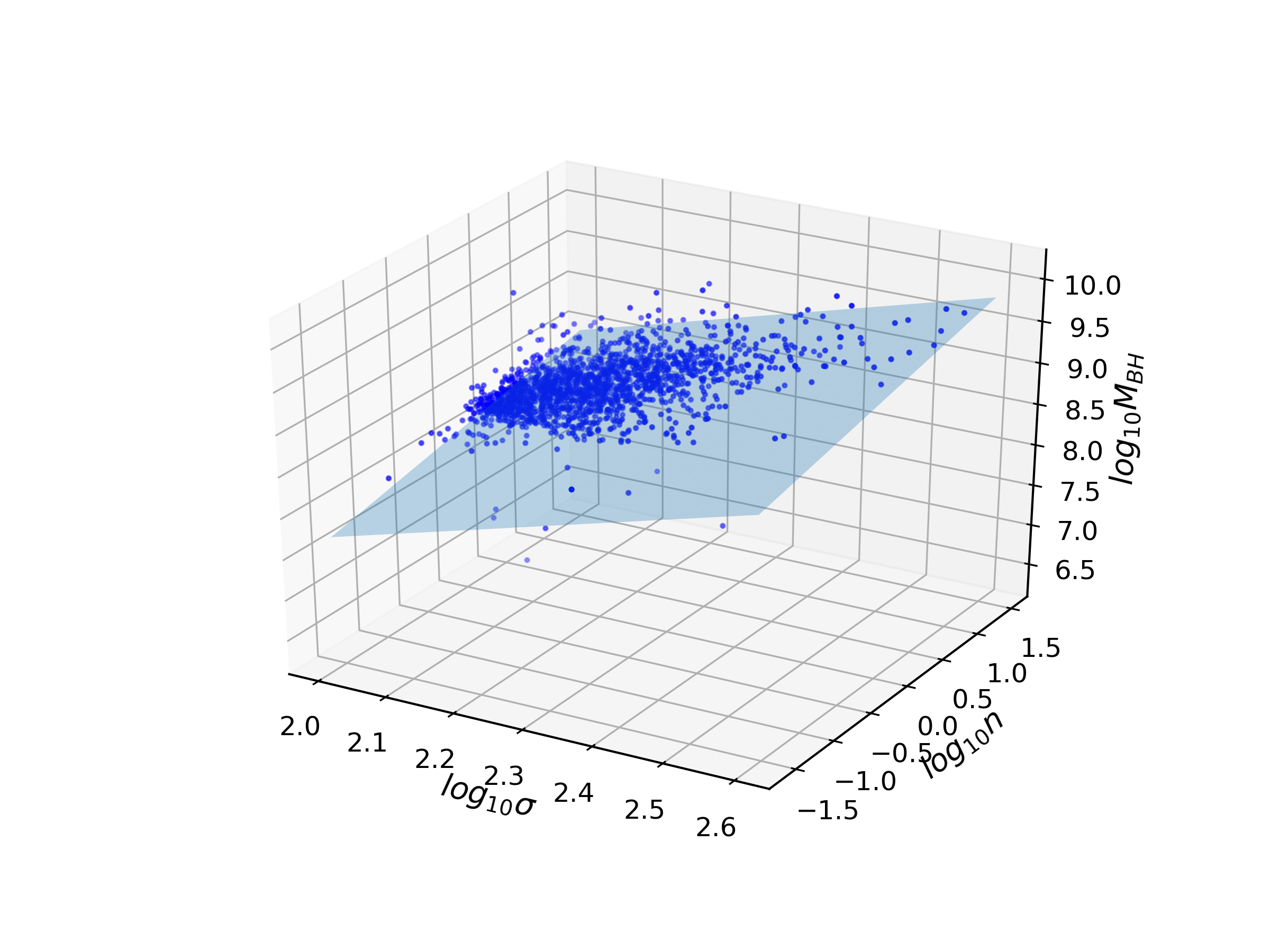} 
\includegraphics[scale=0.5,trim=3cm 1cm 2cm 1cm, clip=true]{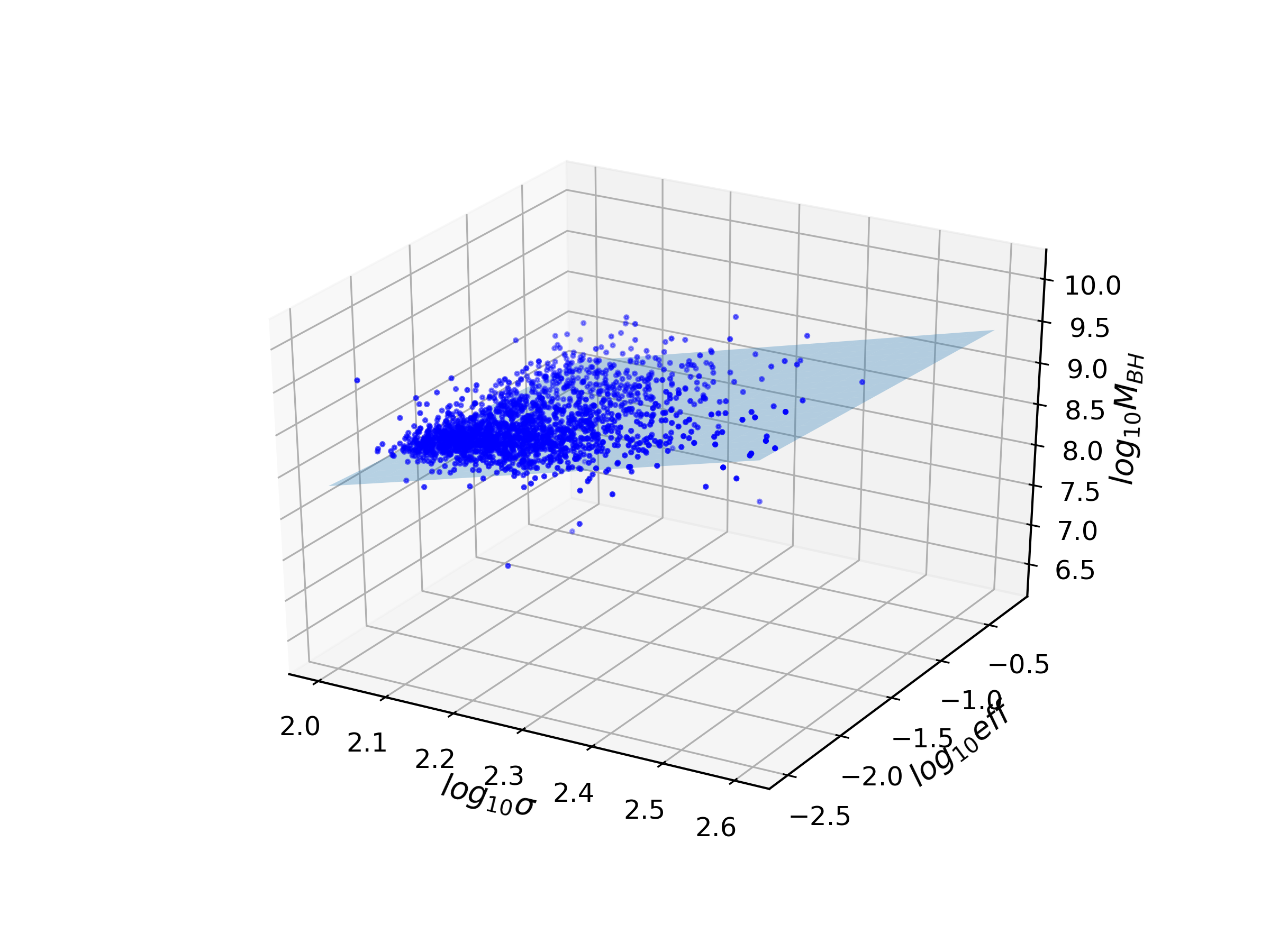}
\caption{SMBH fundamental planes in Illustris (top) and TNG100 (bottom). Left panels show the $M_{BH}-\sigma-n$ planes, where n is the Sersic index, and right panels show the $M_{BH}-\sigma-eff$ planes with $eff$ being the baryon conversion efficiency.
\label{fig:FP}}
\end{center}
\end{figure*}

As is shown in Section~\ref{sec:results2}, in both Illustris and TNG100, the residual of the $M_{BH}-\sigma$ relation correlates with the Sersic index $n$ of the host galaxy, and the baryon conversion efficiency $eff$ for galaxies with $\sigma>100\rm km/s$. It has been shown that the scatter in the observed $M_{BH}-\sigma$ relation can be reduced by introducing a third parameter related to the property of the host galaxy \citep{Marconi2003, Barway2007, Graham2008}. In Figure~\ref{fig:FP}, we show the ``fundamental planes" of $M_{BH}-\sigma-n$ and $M_{BH}-\sigma-eff$ for galaxies with $\sigma>100\rm km/s$ in Illustris and TNG100. 

The functional form we use for the least squares fitting is a simple linear relation: $\rm log_{10}M_{BH}(M_\odot)= C[0]log_{10}\sigma(km/s)+C[1]log_{10}Y+C[2]$, where $Y$ is either $n$ or $eff$. There is no obvious reason to assume that these quantities should correlate with each other in this particular functional form, but for simplicity, we use a linear fit. In other words, we assume that these fundamental planes are flat. Table~\ref{table:FP} summarizes the best-fitting parameters and the root-mean-square (rms) residuals, including those for the simple $M_{BH}-\sigma$ relation without a third parameter. In all cases with a third parameter ($n$ or $eff$), the $M_{BH}-\sigma$ relation is further tightened. 

As Figure~\ref{fig:FP} shows, the $M_{BH}-\sigma$ relation can be seen as a 2D projection of a more fundamental plane, but the two ``fundamental planes" shown in Figure~\ref{fig:FP} are likely only projections of higher dimensional correlations. In fact, for both Illustris and TNG100, the smallest rms residuals are achieved when we include both Sersic index $n$ and $eff$ in the fitting. As is discussed in Section~\ref{sec:results2b}, $M_{BH}$ likely correlates with multiple parameters related to the host galaxy properties, and we have only explored a fraction of them in this work. 

\begin{table}[]
\caption{}
\label{table:FP}
\begin{center}
\begin{tabular}{|c|c|c|c|c|c|}
\hline
            relation                       & simulation & C[0] & C[1] & C[2] & rms residual \\ \hline
\multirow{2}{*}{$M_{BH}-\sigma$}   & Illustris  &  3.69    &  ---    &   0.56   &    0.39      \\ \cline{2-6} 
                                   & TNG100        &  2.40   &  ---   &  3.41   &   0.22       \\ \hline
\multirow{2}{*}{$M_{BH}-\sigma-n$} & Illustris  &  2.62    &  0.65    &   2.68   &     0.36     \\ \cline{2-6} 
                                   & TNG100        &   2.28    &     0.23  &    3.54   &      0.20    \\ \hline
\multirow{2}{*}{$M_{BH}-\sigma-eff$} & Illustris  &   3.51   &   -0.21   &   0.63   &   0.38       \\ \cline{2-6} 
                                   & TNG100        &    2.28  &  -0.12    &   3.47   &   0.21       \\ \hline
\end{tabular}\\ \ \\
Parameters of $M_{BH}-\sigma$ relation and SMBH ``fundamental planes'' in Illustris and TNG100. In all cases, adding a third parameter ($n$ or $eff$) reduces the rms residual of the $M_{BH}-\sigma$ relation. 
\end{center}
\end{table}

\end{document}